\documentclass[small]{acmart}

\usepackage{booktabs} 
\usepackage{enumitem}
\usepackage{makecell}
\usepackage{textcomp}
\usepackage{amsmath}
\usepackage{algorithm} 
\usepackage[noend]{algpseudocode}
\usepackage{pdfpages}
\usepackage{graphicx}
\usepackage{mathrsfs}
\usepackage{xcolor}
\usepackage{epsfig}
\usepackage{graphicx}
\usepackage{multirow}
\DeclareMathOperator*{\argmax}{arg\,max}

\usepackage{enumerate}
\usepackage{balance}
\usepackage[normalem]{ulem}
\usepackage{subfigure}

\usepackage{lineno}

\usepackage{xcolor,colortbl}
\definecolor{mygray}{gray}{0.88}
\usepackage[normalem]{ulem} 

\AtBeginDocument{%
  \providecommand\BibTeX{{%
    \normalfont B\kern-0.5em{\scshape i\kern-0.25em b}\kern-0.8em\TeX}}}

\setcopyright{acmlicensed}
\acmJournal{TOIS}
\acmYear{2025} \acmVolume{1} \acmNumber{1} \acmArticle{1} \acmMonth{1}\acmDOI{10.1145/3717831}

\begin{document}

\title{Vague Preference Policy Learning for Conversational Recommendation}


\author{Gangyi Zhang}
\affiliation{%
  \institution{University of Science and Technology of China}
  \streetaddress{100 Fuxing Road}
  \city{Hefei}
  \country{China}
}
\email{gangyi.zhang@outlook.com}

\author{Chongming Gao}
\authornote{Corresponding author.}
\affiliation{%
  \institution{University of Science and Technology of China}
  \streetaddress{100 Fuxing Road}
  \city{Hefei}
  \country{China}
}
\email{chongming.gao@gmail.com}

\author{Wenqiang Lei}
\affiliation{%
  \institution{Sichuan University}
  \city{Chengdu}
  \country{China}
}

\author{Xiaojie Guo}
\affiliation{%
  \institution{IBM Thomas.J. Watson Research Center}
  \country{USA}
}

\author{Shijun Li}
\affiliation{%
  \institution{The University of Texas at Austin}
  \country{USA}
}

\author{Hongshen Chen}
\affiliation{%
  \institution{JD. com Inc}
  \city{Chengdu}
  \country{China}
}

\author{Zhuozhi Ding}
\affiliation{%
  \institution{JD. com Inc}
  \city{Chengdu}
  \country{China}
}
\author{Sulong Xu}
\affiliation{%
  \institution{JD. com Inc}
  \city{Chengdu}
  \country{China}
}

\author{Lingfei Wu}
\affiliation{%
  \institution{Anytime.AI}
  \country{USA}
}

\renewcommand{\shortauthors}{Zhang, et al.}

\begin{abstract}

Conversational recommendation systems (CRS) effectively address information asymmetry by dynamically eliciting user preferences through multi-turn interactions. However, existing CRS methods commonly assume that users have clear, definite preferences for one or multiple target items. This assumption can lead to over-trusting user feedback, treating accepts/rejects as definitive signals to filter items and reduce the candidate space, potentially causing over-filtering and excluding relevant alternatives.
In reality, users often exhibit vague preferences, lacking well-defined inclinations for certain attribute types (e.g., color, pattern), and their decision-making process during interactions is rarely binary. Instead, users' choices are relative, reflecting a range of preferences rather than strict likes or dislikes. 
To address this issue, we introduce a novel scenario called Vague Preference Multi-round Conversational Recommendation (VPMCR), which employs a soft estimation mechanism to assign non-zero confidence scores to all candidate items, accommodating users' vague and dynamic preferences while mitigating over-filtering.

In the VPMCR setting, we introduce a solution called Vague Preference Policy Learning (VPPL), which consists of two main components: Ambiguity-aware Soft Estimation (ASE) and Dynamism-aware Policy Learning (DPL). ASE aims to accommodate the ambiguity in user preferences by estimating preference scores for both directed and inferred preferences, employing a choice-based approach and a time-aware preference decay strategy. DPL implements a policy learning framework, leveraging the preference distribution from ASE, to guide the conversation and adapt to changes in users' preferences for making recommendations or querying attributes.

Extensive experiments conducted on diverse datasets demonstrate the effectiveness of VPPL within the VPMCR framework, outperforming existing methods and setting a new benchmark for CRS research. Our work represents a significant advancement in accommodating the inherent ambiguity and relative decision-making processes exhibited by users, improving the overall performance and applicability of CRS in real-world settings.
\end{abstract}



\maketitle
\begin{figure*}[t]
    \centering
	   \includegraphics[width=1\linewidth]{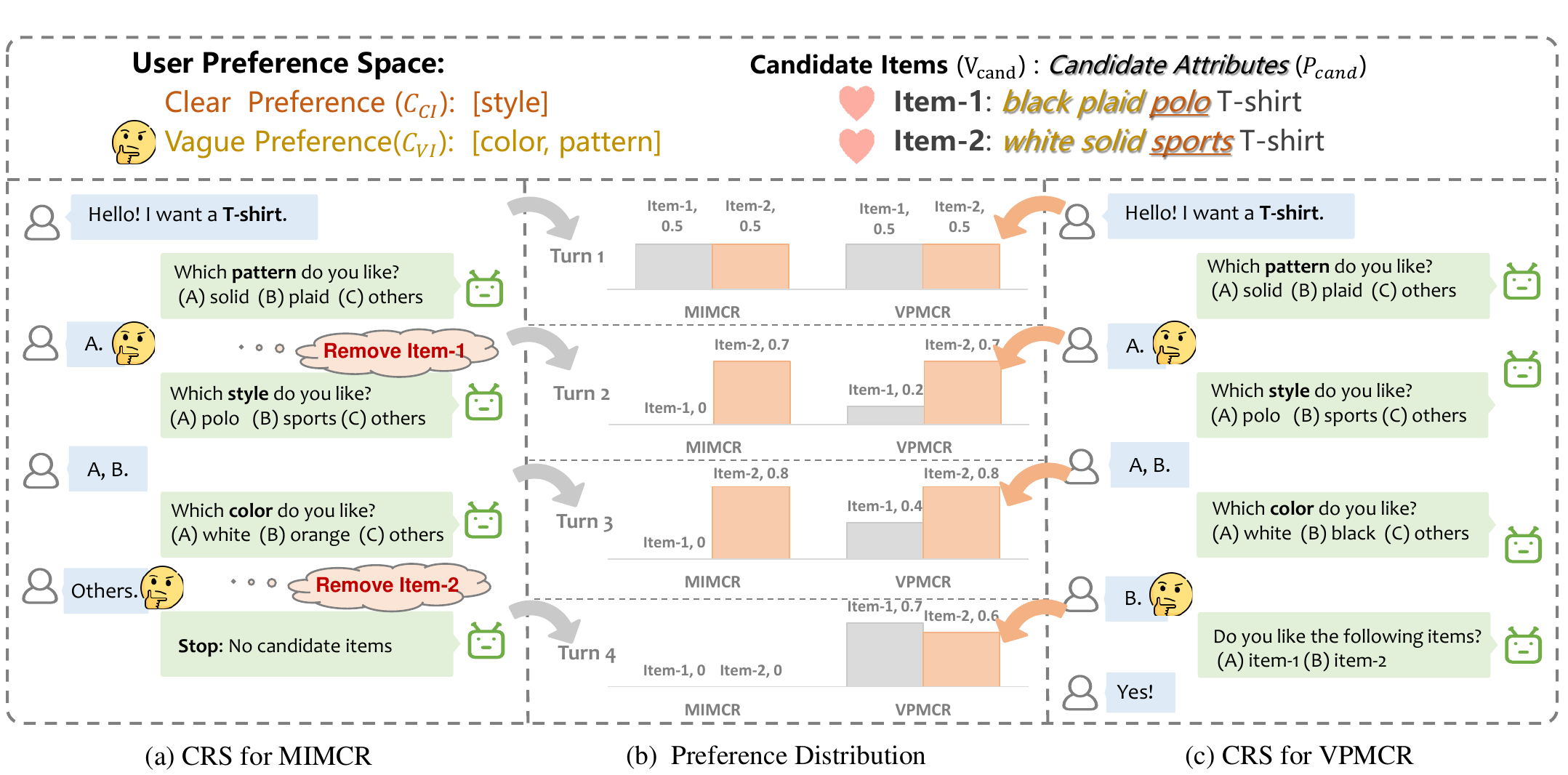}
    \caption{This provides a simple illustration contrasting user preference modeling under the MIMCR and VPMCR scenarios. In the MIMCR scenario (Figure (a)), non-clicking attributes may lead to the premature removal of potential target items, causing a sudden and possibly erroneous narrowing of the user's preference distribution, as depicted in the left of Figure (b). In the VPMCR scenario (Figure (c)), both clicking and non-clicking attributes contribute to the evolution of a soft preference distribution across the entire item space, able to accommodate vague or dynamic user preferences. Unlike MIMCR, under VPMCR the potential for recommending preference items is retained as shown on the right side of Figure (b).  
    }
    \label{fig:intro-exa}
\end{figure*}

\section{Introduction}
\label{sec:introduction}

Conversational recommendation systems (CRS) have drawn a lot of research attention recently. These systems interact with users to elicit preferences, understand motivations, and address the long-standing information asymmetry problem \cite{gao2021advances}. Despite a lot of efforts having been made, CRS is still far from mature and there is a long way to go in developing highly intelligent agents that can communicate with users like real humans. In this regard, some researchers narrow down their efforts to a few scenarios ~\cite{Sun:2018:CRS:3209978.3210002, lei20estimation, zhang2022multiple} where specific problems are highlighted and solved. 

One widely adopted scenario \cite{lei20estimation,lei2020interactive,xu2021adapting} is \textbf{Multi-round Conversational Recommendation (MCR)}, where the system can ask for attributes or make recommendations multiple times, and the user accepts or rejects accordingly. Multi-round Conversational Recommendation (MCR) scenario is a widely adopted setting since it is one of the most straightforward and effective scenarios in CRS \cite{lei20estimation,lei2020interactive,xu2021adapting}. Traditionally, MCR assumes that users have a single preferred item in mind throughout the conversation. During these interactions, when the system queries an attribute instance (e.g., “Do you like the black color?”), users respond with binary answers, either affirming (clicking “Yes”) or rejecting the option. In this setting, the user is assumed to have a clear preference space anchored on the attributes of their target item. For instance, if the target item is a black plaid polo T-shirt, the user is expected to affirm attributes like "black," "plaid," and "polo" when queried, while rejecting attributes that do not match their target item.

Recognizing the limitation of MCR in accommodating users with multiple preferred items, the \textbf{ Multi-Interest Multi-round Conversational Recommendation (MIMCR) scenario} \cite{zhang2022multiple} was introduced. MIMCR extends MCR by allowing users to accept multiple attribute instances, thereby catering to users with interests spanning multiple target items. For example, if a user likes both a black plaid polo T-shirt and a white solid sports T-shirt, his preference space encompasses the attributes of both items. MIMCR also employs a hierarchical questioning strategy, where the system first selects an attribute type (e.g., color) and then queries about multiple instances within that type (e.g., white or black), making the interaction more efficient.

While MCR and MIMCR have their merits, they make two key simplifying assumptions: (1) users have only clear preferences, and (2) users' decision-making during the interaction is a binary process of accepting or rejecting options. Regarding the first assumption, a clear preference refers to a user's definite and singular inclination towards a specific attribute instance or type. However, in real-world scenarios, users often exhibit vague preferences for certain attribute types (e.g., color, pattern), indicating that they do not have a well-defined preference but rather a range of acceptable options. 
For instance, a user may have a clear preference for a T-shirt style (e.g., "plaid") but a vague preference for colors, being open to various options. Moreover, due to these vague preferences, users' decision-making is rarely a binary process \cite{bettman1998constructive}. Their choices are often relative rather than absolute, reflecting a range of preferences rather than a strict like or dislike. Consequently, users' behavior during the interaction does not always represent a straightforward binary signal (acceptance or rejection) but rather a degree of preference tendency. We cannot interpret user feedback (such as clicking an option) as a definitive acceptance or rejection; instead, it should be viewed as an explicit expression of directed preference or an implicit indication of an inferred preference based on their behavior (e.g., not clicking an option).

In scenarios where users exhibit vague or dynamic preferences, the MCR or MIMCR may fall short in addressing this ambiguity. These models commonly interpret user feedback as a clear indicator for filtering candidate items, which can lead to excessive filtering and the inadvertent exclusion of pertinent alternatives. This issue is exacerbated as numerous potential items are eliminated following the user's selection or non-selection of specific attributes.

Let us consider a toy example in Figure \ref{fig:intro-exa}, where a user wants to buy a T-shirt and has a vague preference for color and pattern. Suppose the CRS asks the user “Which pattern do you like?” and displays three options: “(A) solid, (B) plaid, and (C) others”. If the user clicks on option (A), existing CRS methods, such as Multi-Interest Multi-round Conversational Recommendation (MIMCR), assume that the user has clear preference for ‘solid’, while non-clicking attributes (e.g., “plaid“) are unpreferred, leading to the removal of potential target items (e.g., “item-1“). This may cause a sudden and possibly inaccurate narrowing of the user’s preference distribution, as depicted in Figure \ref{fig:intro-exa} (b). This inference may affect the decision reasoning of the subsequent conversation, resulting in the omission of relevant attributes (e.g., the “black” color of “item-1” that may be preferred by the user was not displayed in the third turn). Then, when the user does not click the attribute “white” of “item-2” due to their vague preference, item-2 is removed again, resulting in a recommendation mismatch or dissatisfaction.

To address these limitations, we introduce the Vague Preference Multi-round Conversational Recommendation (VPMCR) scenario. VPMCR employs a soft estimation mechanism that assigns non-zero confidence scores to all candidate items, thereby accommodating users' vague or dynamic preferences and avoiding the over-filtering problem of existing CRS methods. For instance, in Figure \ref{fig:intro-exa} (c) , when the user does not click the "plaid" option, VPMCR does not necessarily interpret this as a complete dislike for plaid patterns. Instead, it maintains a certain level of preference for "plaid" items, capturing the user's implicit vague preferences. This soft estimation approach allows VPMCR to model the evolution of preference distributions across the entire item space, as shown in Figure \ref{fig:intro-exa} (b),  mitigating the rigid filtering characteristic of MIMCR and MCR while maintaining diversity and accuracy in the recommendations.

Addressing the VPMCR scenario poses several challenges, including accurately estimating the vagueness of user preferences and modeling the relative nature of users' decision-making processes during the conversation. To effectively navigate VPMCR scenario, we introduce the \textbf{Vague Preference Policy Learning (VPPL)} framework, comprising two main components:

1. \textbf{Ambiguity-aware Soft Estimation (ASE)}: ASE aims to accommodate the ambiguity of users' preferences by estimating preference scores for both directed and inferred preferences. It employs a choice-based approach to extract preferences from user feedback , where directed preferences are derived from the options explicitly clicked by the user, while inferred preferences are deduced from the non-clicked options based on the user's behavior (e.g., not clicking an option does not necessarily indicate a complete dislike). To capture users' dynamic preferences, ASE employs a time-aware preference decay strategy, which gives more weight to recent preferences while gradually diminishing the influence of historical preferences.

2. \textbf{Dynamism-aware Policy Learning (DPL)}: DPL implements a policy learning framework, leveraging the preference distribution from ASE, to guide the conversation. It employs a dynamic heterogeneous graph to model the conversation, with edge weights reflecting ASE's soft estimation scores. To enhance graph modeling and policy learning efficiency, we have introduced a novel graph sampling strategy and preference-guided action pruning technique. The policy is optimized with reinforcement learning algorithm to predict the best action based on conversation representation.

In summary, this paper makes the following significant contributions to the field of Conversational Recommendation Systems (CRS):
\begin{itemize}
\item  We critically analyze the limitations of current CRS frameworks, particularly in handling user preferences that are inherently vague and dynamic. To address this, we introduce the Vague Preference Multi-round Conversational Recommendation (VPMCR) scenario , which more accurately reflects the ambiguity and relative decision-making processes exhibited by users in real-world CRS environments.
\item  We develop the Vague Preference Policy Learning (VPPL) framework for the VPMCR scenario. VPPL employs a unified approach that estimates users' vague preferences and models their relative decision-making processes during the conversation. This framework represents a significant advancement in accommodating the complexities of user preferences and decision-making in CRS.
\item  Through extensive experimental analysis conducted on four diverse real-world datasets, we demonstrate the superior performance of VPPL within the VPMCR framework. These experiments underscore the practical applicability and effectiveness of our proposed solution, setting a new benchmark for future research in CRS.
\end{itemize}

\section{Related Work}\label{sec:related}
In this section, we provide an overview of the existing research in the domains of conversational recommendation systems, reinforcement learning-based recommendation, and graph-based recommendation. We highlight key developments and contributions in each of these areas, setting the context for our work.

\subsection{Conversational recommendation system} (CRSs) is a novel solution to recommendation that leverage natural language to effectively elicit dynamic user preferences that align with their real needs through multiple rounds of real-time interaction. CRS is considered to be a cutting-edge discipline that incorporates dialogue systems, recommendation systems, and interactive systems \cite{gao2021advances}.

According to the focus on different functions and settings, existing CSR methods can be roughly divided into two types: dialogue-based recommendation \cite{nips18/DeepConv,zhou2020topicguided,chen-etal-2019-towards,kdd2022UniCRS,wu2019proactive,tois2022unimind} and multi-round conversational recommendation (MCR) \cite{lei2020interactive,deng2021unified,xu2021adapting,10.1145/3523227.3546755,gao2022kuairec,li2020seamlessly,tkde2023Long-shortCRS, cikm24_TCRS}. 
Our study concentrates on the latter, MCR, which is hailed as the most realistic CRS setting. 

Dialogue-based recommenders rely on pre-collected dialogue corpora to generate responses consistent with the predefined corpus, prioritizing text generation consistency over efficient interaction and recommendation strategies.
For instance, CFCRS \cite{wang2023improving} uses counterfactual data augmentation to simulate user preference changes and generate diverse dialogue data.
In recent advancements, Large Language Models (LLMs) have begun to play a crucial role in enhancing conversational recommendation systems (CRS) by providing more contextually relevant interactions. ~\citet{zhang2023user} and ~\citet{friedman2023leveraging}, have explored LLMs' potential to dynamically adapt to user preferences within a user-centric conversational framework. However, these studies often provide a feasible solution framework without concrete solutions, and LLM-based methods still face challenges in reliable evaluation\cite{yang2024behavior}. 
Unlike dialogue-based recommenders that need to extract information or generate responses through raw natural language \cite{kdd2022UniCRS}, MCR focuses on the core logic of the interaction strategy which involves asking questions \cite{10.1145/3397271.3401180,10.1145/3397271.3401180,10.1145/3477495.3532077, www2022hierarchicalCRS, lin2023enhancing, jin2023lending} and making recommendations.

Traditional MCR models, however, have a limitation: they permit users to select only one preferred attribute value at a time, constraining the expression of user preferences during interactions. To address this, \citet{zhang2022multiple} proposed the Multi-Interest Multi-round Conversational Recommendation (MIMCR) setting, allowing users to choose multiple attribute options. While this approach marked a significant advancement, it adhered to MCR's underlying philosophy of filtering out unmentioned items, which can be problematic as users may not always have precise preferences. HutCRS \cite{emnlp2023HutCRS} enhances the MIMCR setting by introduce a hierarchical interest tree, enhancing conversational efficiency and user preference targeting.

To further refine the MIMCR approach, some innovations works \cite{chu2023meta, tois2023FuzzyCRS, cikm24_TCRS} have been instrumental. MetaCRS \cite{chu2023meta} proposes to overcome the unrealistic assumption of existing conversational recommendation works by learning a meta policy that can adapt to new users with only a few trials of conversational recommendations. An additional progression involves the incorporation of explicit textual feedback within user simulations, as elucidated in \citet{tois2023FuzzyCRS}. This approach, simulating real-world user uncertainties with responses such as "I don't know," highlights the need for CRS to interpret and navigate vague or uncertain feedback. Despite its relevance, explicit textual feedback like this is not commonly found in typical user interactions. 

Our Vague Preference Multi-round Conversational Recommendation (VPMCR) setting is designed to address these nuances. By focusing on implicit vague preferences, VPMCR effectively captures the intricacies of user decision-making processes, which are often marked by indecisiveness or evolving preferences. This approach not only aligns the recommendation process more closely with real-world scenarios but also yields more accurate and satisfying recommendations by understanding and adapting to the subtleties of user needs.

\subsection{RL-based Recommendation} Reinforcement Learning (RL) is a type of Machine Learning. It considers how an agent (e.g., a machine) should automatically make decisions within a specific context to pursue a long-term goal. The agent learns and adjusts its policy based on the reward feedback (i.e., reinforcement signals) given by the environment. Recently, RL has shown its effectiveness in recommendation \cite{afsar2022reinforcement,deffayet2023offline,gao2023alleviating,RPP_TOIS,gao2024sprec}. As fitting user interest is not a bottleneck for now, recommenders care more about users' long-term satisfaction \cite{ResAct,10.1145/3534678.3539040,wang2022best}. For instance, \citet{10.1145/3488560.3498526} use RL to generate the proper questions that can maximally make the system help users search desired products. \citet{gao2022cirs} integrate causal inference into offline RL to maximize users' long-term satisfaction by removing filter bubbles. \citet{10.1145/3534678.3539095} propose an RL-based dispatching solution for ride-hailing platforms that can conduct robust and efficient on-policy learning and inference while being adaptable for full-scale deployment. In this work, we use RL to learn a policy that can automate question-asking and item recommendation.

\subsection{Graph-based Recommendation} Graph-based recommender systems have drawn a lot of research attention \cite{10.1145/3534678.3539452,10.1145/3534678.3539458,10.1145/3447548.3467384, wu2022state}. By arranging the various entities (e.g., users, items, and attributes) in a heterogeneous graph, we can leverage lots of properties in modeling the collaborative signals. In CRS, the knowledge graph is utilized in enriching the system with additional knowledge \cite{lei2020interactive,xu2020user,zhou2020improving,moon-2019-opendialkg,zhou2020interactive,mao2024invariant}. For example, to better understand concepts that a user mentioned, \citet{zhou2020improving} propose to incorporate two external knowledge graphs (KGs): a word-oriented KG providing relations (e.g., synonyms, antonyms, or co-occurrence) between words and an item-oriented KG carrying structured facts regarding the attributes of items. With the increasing in nodes, the computational overhead is too large to satisfy the requirement for real-time interaction. Hence, we propose a pruning strategy to overcome this work.

\section{Problem Definition} \label{sec:definition}
\textbf{Vague Preference Multi-round Conversational Recommendation (VPMCR).}
In the VPMCR scenario, we consider a dynamic conversation between a user and a conversational recommendation system (CRS). The user has a clear preference space, denoted as $\mathcal{C}_{CI}$ (e.g., "style" in Figure \ref{fig:intro-exa}), and a vague preference space, denoted as $\mathcal{C}_{VI}$ (e.g., "color" and "pattern" in Figure \ref{fig:intro-exa}).

The conversation begins with the user specifying a query attribute $p_0$ (e.g., "T-shirt"), which initializes the candidate item set $\mathcal{V}_{cand}$  containing all relevant items (e.g., all "T-shirts") and the candidate attribute set $\mathcal{P}_{cand}$ containing all attributes of those items. 

During the conversation, the CRS can either ask questions about attributes or provide recommendations. When the CRS asks questions, the user responds accordingly with their behavior depending on whether the attribute type $c$ belongs to their clear or vague preference space. If $c \in \mathcal{C}_{CI}$, the user \emph{honestly} accepts or rejects the displayed attributes. However, if $c \in \mathcal{C}_{VI}$, the user may \emph{randomly} accept or reject a potentially preferred attribute.
When the CRS provides recommendations, the user can accept or reject one or more items from the recommended set $\mathcal{V}_{rec}$. 

The conversation proceeds through multiple iterations of the CRS asking/recommending and the user responding, until a successful recommendation is made or the maximum number of turns is reached. The VPMCR scenario differs from previous MCR or MIMCR settings in that it does not filter $\mathcal{V}_{cand}$ based on the user's clicking or non-clicking attributes. Instead, it only removes $\mathcal{V}_{rec}$ from $\mathcal{V}_{cand}$ when the recommendation fails. Additionally, all candidate attributes linked to candidate items are maintained in $\mathcal{P}_{cand}$.

\section{Notations and Symbols}
\label{sec:notations}

To assist the reader in understanding the symbols and variables used throughout the article, we provide a comprehensive list of notations in Table \ref{tab:notations}.

\begin{table}[H]
\centering
\begin{tabular}{|c|l|}
\hline
\textbf{Symbol} & \textbf{Description} \\ \hline
$u$ & User. \\ \hline
$v$ & Item. \\ \hline
$p$ & Attribute associated with items (e.g., "black color", "polo style"). \\ \hline
$\mathcal{C}_{CI}$ & Clear preference space, representing user preferences that are explicit and \\
& predefined in the user simulator to reflect concrete preferences for certain attributes. \\ \hline
$\mathcal{C}_{VI}$ & Vague preference space, representing user preferences that are uncertain or flexible, \\
& defined in the user simulator to simulate real-world scenarios of uncertainty. \\ \hline
$p_0$ & Initial query attribute provided by the user, e.g., "T-shirt". \\ \hline
$\mathcal{V}_{cand}$ & Candidate item set containing all items relevant to the user's query at the current stage. \\ \hline
$\mathcal{V}_{rec}$ & Set of items recommended by the CRS during a specific interaction turn. \\ \hline
$\mathcal{P}_{cand}$ & Candidate attribute set associated with the items in $\mathcal{V}_{cand}$. \\ \hline
$\mathcal{P}_{click}^{(t)}$ & Set of attributes explicitly clicked by the user in turn $t$. \\ \hline
$\mathcal{P}_{noclick}^{(t)}$ & Set of attributes presented but not clicked by the user in turn $t$. \\ \hline
$\mathcal{P}_{noshow}^{(t)}$ & Set of attribute instances not displayed to the user in turn $t$. \\ \hline
$\mathcal{V}_{top-N}^{(t)}$ & Set of top-N candidate items selected for potential recommendation in turn $t$, \\
 & based on the user's estimated preference scores, used within the user simulator. \\ \hline
$\mathcal{P}_{top-N}^{(t)}$ & Set of top-N candidate attributes selected for querying, determined by estimated preference. \\ \hline
$\mathcal{A}_{\text{action}}^{(t)}$ & Pruned action space for turn $t$, consisting of top-N items and attributes \\
 & identified as most relevant to the user's current preferences. \\ \hline
$e_u$ & Embedding vector representing user $u$. \\ \hline
$e_v$ & Embedding vector representing item $v$. \\ \hline
$e_p$ & Embedding vector representing attribute $p$. \\ \hline
$\mathcal{G}_u^{(t)}$ & Dynamic undirected graph representing the conversation state at turn $t$. \\ \hline
$\mathcal{N}^{(t)}$ & Node set in the dynamic graph $\mathcal{G}_u^{(t)}$, including user, items, and attributes. \\ \hline
$\mathbf{A}^{(t)}$ & Weighted adjacency matrix of graph $\mathcal{G}_u^{(t)}$ at turn $t$. \\ \hline
$\mathbf{s}_{conv}^{(t)}$ & Conversation state representation at turn $t$, derived from graph-based modeling. \\ \hline
$Q(s, a; \theta)$ & Q-value function for state-action pairs, approximated using parameters $\theta$. \\ \hline
\end{tabular}
\caption{Summary of Notations and Symbols}
\label{tab:notations}
\end{table}




\begin{figure*}[t]
    \centering
	\includegraphics[width=\linewidth]{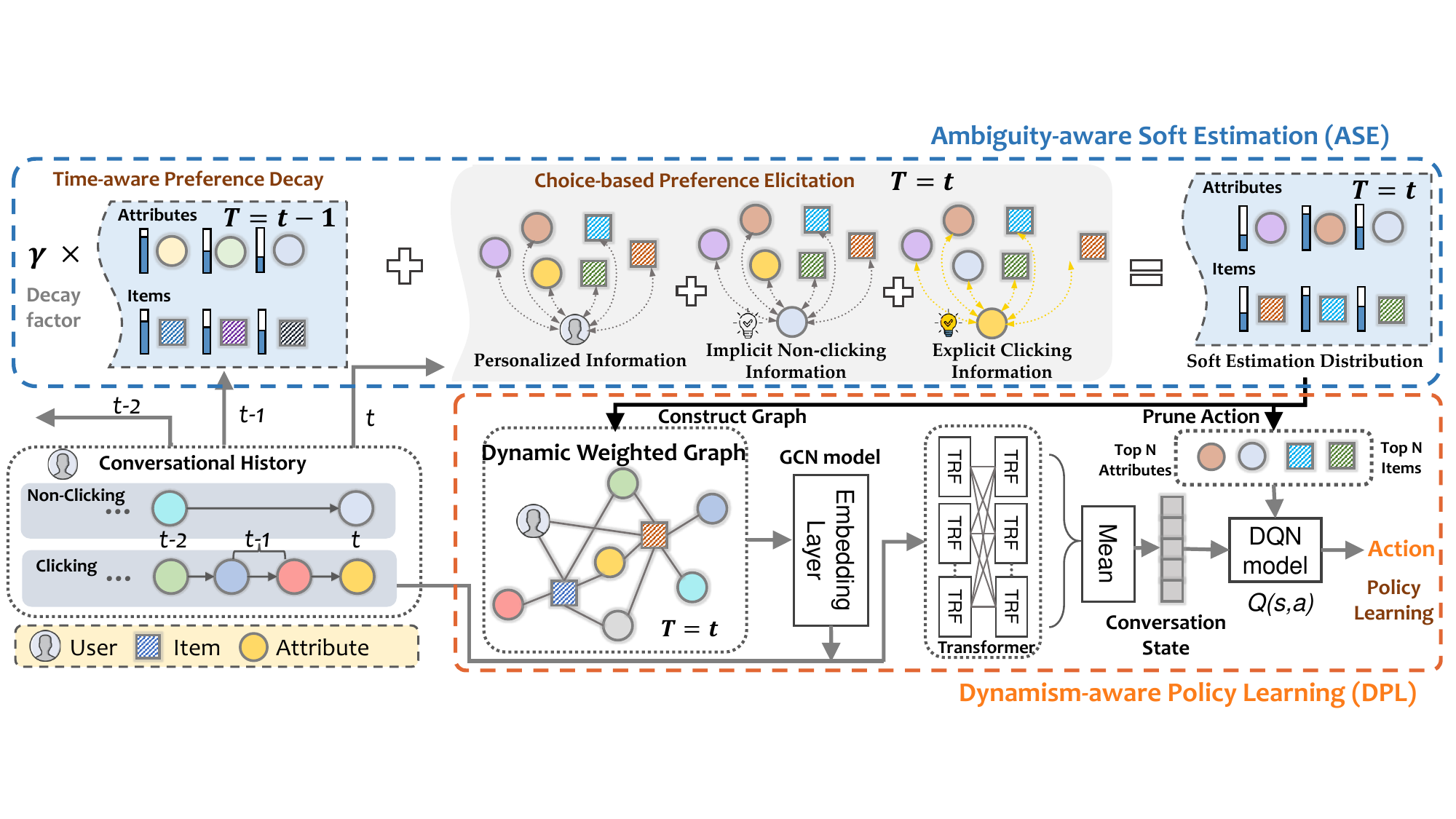}
    \caption{Vague Preference Policy Learning (VPPL) solution for VPMCR scenario. The model comprises two modules: Ambiguity-aware Soft Estimation (ASE) and Dynamism-aware Policy Learning (DPL). ASE models user’s vague preferences and preference decay during the conversation. DPL uses ASE’s output to construct a dynamic graph for conversation state representation and prunes the action space for efficient policy learning. The goal is to adapt to the user’s vague or dynamic preferences.}
    \label{fig:method-exa}
\end{figure*}
\section{METHODOLOGY}
In the VPMCR setting, we introduce a solution called Vague Preference Policy Learning (VPPL), which consists of two main components: Ambiguity-aware Soft Estimation (ASE) and Dynamism-aware Policy Learning (DPL). ASE aims to accommodate the ambiguity in user preferences by estimating preference scores for both directed and inferred preferences, employing a choice-based approach and a time-aware preference decay strategy. DPL implements a policy learning framework, leveraging the preference distribution from ASE, to guide the conversation and adapt to changes in users’ preferences for making recommendations or querying attributes. The overall framework of VPPL is illustrated in Figure \ref{fig:method-exa}, where ASE captures both explicit and implicit feedback to model user preferences, while DPL optimizes decision-making through reinforcement learning-based policy adaptation.

\subsection{Ambiguity-aware Soft Estimation} 

\label{method: ASE}

ASE is engineered to accommodate the inherent uncertainty in user preferences by calculating preference scores based on \emph{directed preferences} (from clicking behavior) and \emph{inferred preferences} (from non-clicking behavior). This is achieved through a refined choice-based approach that also incorporates a time-aware preference decay strategy to dynamically adjust to changing user preferences over the conversation's course.

\subsubsection{Preference Extraction with Choice-based Approach} 
\label{method: ASE-preference-extraction}

In each interaction turn, user preference can be divided into personalized user preference and choice-based preference. Personalized user preference reflects the user’s static or long-term preference for items, which can be learned from historical interaction data using collaborative filtering models or graph-based models. Choice-based preference reflects the user’s dynamic or short-term preference for items and attributes, which can be inferred from the user’s feedback in the current conversation. 

we use embeddings learned from graph methods as the initial representations for users and items. The static preference of user $u$ for item $v$ is represented as:
\begin{equation}
\label{eq:ui_score}
w_{v\mbox{-}u} = e_{u}^{\top} e_v,
\end{equation}
where $e_{u}$ and $e_v$ denote the embedding vectors of user $u$ and item $v$, respectively.

Moving beyond traditional binary feedback models that infer clear and consistent preferences, our choice-based preference extraction method acknowledges the complexity and fluidity of user preferences in conversational recommendation settings. This approach identifies \emph{directed preferences} when users actively choose  attributes, and \emph{inferred preferences} when non-selection suggests a potential yet unstated preference. We evaluate these preferences separately, reflecting the decision-making processes.

To account for both directed and inferred preferences, ASE separately assesses the impact of attributes based on user interactions. For any item \( v \) during turn \( t \), the \emph{directed preferences} from clicked attributes \( \mathcal{P}_{\text{click}}^{(t)} \) and the \emph{inferred preferences} from non-clicked attributes \( \mathcal{P}_{\text{noclick}}^{(t)} \) are evaluated as follows:
\begin{equation}
\label{eq:click_score}
\begin{split}
w_{v\mbox{-}click}^{(t)} = \frac{1}{\lvert \mathcal{P}_{\text{click}}^{(t)} \rvert}
\sum_{p \in \mathcal{P}_{\text{click}}^{(t)}}(e_{v}^{\top} e_p - w_{v\mbox{-}avg}^{(t)}),\\
w_{v\mbox{-}noclick}^{(t)} = \frac{1}{\lvert \mathcal{P}_{\text{noclick}}^{(t)} \rvert}
\sum_{p \in \mathcal{P}_{\text{noclick}}^{(t)}}(e_{v}^{\top} e_p - w_{v\mbox{-}avg}^{(t)}),
\end{split}
\end{equation}
where $\lvert \mathcal{P}_{\text{click}} \rvert$ and $\lvert \mathcal{P}_{\text{noclick}} \rvert$ indicates the number of attributes related to clicked attribute instances and non-clicked attribute instances, respectively.
$w_{v\mbox{-}avg}^{(t)}$ measures the average preference offset towards all unshown attribute instances of the queried attribute type and is used to mitigate over-estimation of the system-displayed attribute instances, which is defined as:
\begin{equation}
\label{eq:avg_score}
w_{v\mbox{-}avg}^{(t)} = 
\sum_{p \in \mathcal{P}_{\text{noshow}}^{(t)}} e_{v}^{\top} e_{p} \bigg / 
\lvert \mathcal{P}_{\text{noshow}}^{(t)} \rvert,
\end{equation}
where $e_{v}$ and $e_{p}$ represent the embedding vectors of item $v$ and attribute $p$, respectively, and $\mathcal{P}_{\text{noshow}}^{(t)}$ refers to the set of all unshown attribute instances of the queried attribute type in turn $t$.

By considering both the personalized preferences and the choice-based preference in turn $t$, the users' preference for item $v$ in turn $t$ can be calculated as:
\begin{equation}
\label{eq:item_score}
w_v^{(t)} =\sigma (w_{v\mbox{-}u} + \lambda_{1} w_{v\mbox{-}click}^{(t)} + \lambda_{2} w_{v\mbox{-}noclick}^{(t)}).
\end{equation}
where $\sigma$ is the sigmoid function. $\lambda_{1}$ and $\lambda_{2}$ represent the information intensity coefficients of the information contained in the user's clicked attribute and the user's unclicked attribute, respectively.


This method models users' \emph{directed preferences} and \emph{inferred preferences} separately, reflecting their relative decision-making processes when responding to choice-based questions.

\subsubsection{Time-aware Preference Decay} 
\label{method: ASE-preference-decay}
One of the key challenges in the VPMCR scenario is to capture the user’s dynamic preferences that may change over the course of the conversation. To address this challenge, we propose a time-aware preference decay mechanism that adjusts the influence of historical preferences based on their recency. This mechanism enables the model to focus more on the user’s real-time feedback in the current turn and mitigate the over-emphasized impact of the user’s clicking behavior in previous turns.

The time-aware preference decay mechanism is based on the assumption that the user’s preferences in the current turn are more indicative of their vague preferences than their preferences in earlier turns. Therefore, we assign a decay factor $\gamma$ to each turn, satisfying $0\le \gamma\le 1$, such that the preferences in the current turn have the highest weight and the preferences in the earliest turn have the lowest weight. The decay factor $\gamma$ can be seen as a measure of the user’s preference volatility. 

To combine the user’s current preference with historical decay preferences, we use a linear decay function to estimate the user’s global preference toward each item and attribute. Specifically, for item $v$, the user’s global preference in turn $t$ is calculated as follows:
\begin{equation}
\label{eq:decay_item_score}
w_v^{(t)} = w_v^{(t)} + \gamma w_v^{(t-1)},
\end{equation}
which can be unfolded as:
\begin{equation}
\label{eq:generalized_decay_item_score}
w_v^{(t)} = \sum_{i=0}^{t-1} \gamma^{t-i-1} w_v^{(i)}.
\end{equation}
where $w_v^{(i)}$ denotes the user’s preference for item $v$ in turn $i$, estimated by the choice-based preference extraction method in Section ~\ref{method: ASE-preference-extraction}.

Finally, for turn $t$, the user's global preference distribution for items $f_u^{(t)}(v)$ can be calculated by estimating the user's global preference $w$ for each item $v$ in the candidate item set $\mathcal{V}_{\text{cand}}$. When the size of the candidate item set is $n$, the soft estimation distribution for items is shown as follows:
\begin{equation}
\label{eq:item_distribution}
f_u^{(t)}(v) = \{ \frac{w_{v_1}^{(t)}}{\sum_{i=1}^{n} w_{v_i}^{(t)}}, \frac{w_{v_2}^{(t)}}{\sum_{i=1}^{n} w_{v_i}^{(t)}}, ..., \frac{w_{v_n}^{(t)}}{\sum_{i=1}^{n} w_{v_i}^{(t)}}\}.
\end{equation}

Similarly, by replacing items with attributes in the aforementioned equations, we derive the user's global preference distribution towards the candidate attribute set $\mathcal{P}_{\text{cand}}$. When the size of the candidate attribute set is $m$, the soft estimation for attributes is depicted by the following distribution:
\begin{equation}
\label{eq:attribute_distribution}
f_u^{(t)}(p) = \{ \frac{w_{p_1}^{(t)}}{\sum_{i=1}^{m} w_{p_i}^{(t)}}, \frac{w_{p_2}^{(t)}}{\sum_{i=1}^{m} w_{p_i}^{(t)}}, ..., \frac{w_{p_m}^{(t)}}{\sum_{i=1}^{m} w_{p_i}^{(t)}}\}.
\end{equation}

By incorporating the recency and volatility of user preferences, the time-aware preference decay mechanism can effectively balance the influence of historical preferences and real-time feedback.

\subsection{Dynamism-aware Policy Learning (DPL)}
The Dynamism-aware Policy Learning (DPL) module plays a pivotal role in the Vague Preference Policy Learning (VPPL) framework, primarily by utilizing the preference distributions deduced by the Ambiguity-aware Soft Estimation (ASE) module. DPL strategically guides the conversational trajectory, adeptly adapting to the evolving preference landscape of the user. This module is instrumental in aligning conversational decisions with the user’s current, albeit vague or dynamic, preferences. Crucially, DPL contributes to optimizing the long-term conversational rewards, thereby enhancing the overall efficacy of Conversational Recommendation Systems (CRS) for users with fluid preferences. In achieving these objectives, DPL employs innovative techniques, including a novel graph sampling strategy and a preference-guided action pruning mechanism. These methodologies enable DPL to efficiently navigate the complex conversational pathways dictated by ASE’s nuanced preference assessments, ensuring a more responsive and adaptive CRS experience.

\subsubsection{Graph-based Conversation Modeling}
\label{method: cur_conv_state}
In the Graph-based Conversation Modeling section, we build on previous work ~\cite{deng2021unified, zhang2022multiple} to represent the current conversation state at turn $t$ using a dynamic undirected graph $\mathcal{G}_u^{(t)} = (\mathcal{N}^{(t)}, \mathbf{A}^{(t)})$. 
This graph is a subgraph of the heterogeneous graph, which consists of users, items, and attributes. 

The nodes in the graph, $\mathcal{N}^{(t)}$, are defined as follows:
\begin{equation}
\label{eq:graph_node}
\mathcal{N}^{(t)}=\{u\} \cup \mathcal{P}_{\text {click}} \cup \mathcal{P}_{n\mbox{-}\text {click}} \cup\mathcal{P}_{\text {cand}}^{(t)} \cup \mathcal{V}_{\text {sample}}^{(t)}
\end{equation}
The node set $\mathcal{N}^{(t)}$ contains the user $u$, user's historical clicked attributes $\mathcal{P}_{\text {click}}$, user's historical non-clicked attributes $\mathcal{P}_{n\mbox{-}\text {click}}$, current candidate attributes $\mathcal{P}_{\text {cand}}^{(t)}$, and current sampled candidate items $\mathcal{V}_{\text {sample}}^{(t)}$. 

In the VPMCR setting, every item is assigned a probability value of being recommended. Thus, there may be numerous candidate items, which may cause the dynamic graph too large to meet the computational requirement of the real-time interaction.

To address the computational challenges posed by a large number of candidate items, we introduce a sampling strategy for $\mathcal{V}_{\text{sample}}^{(t)}$. This strategy involves randomly selecting a subset of candidate items at each turn, ensuring manageable computational complexity while maintaining a broad representation of possible recommendations.

The weighted adjacency matrix, $\mathbf{A}^{(t)}$, is defined as:
\begin{equation}
\label{eq:graph_matrix}
\begin{array}{l}
A_{i, j}^{(t)}=\left\{\begin{array}{ll}
w_{v}^{(t)}, & \text { if } n_{i}=u, n_{j} \in \mathcal{V} \\
1, & \text { if } n_{i} \in \mathcal{V}, n_{j} \in \mathcal{P} \\
0, & \text { otherwise }
\end{array}\right.
\end{array}
\end{equation}
The weight $w_v^{(t)}$ denotes the user's estimated vague preference for the item $v$, which is calculated via Eq.~(\ref{eq:generalized_decay_item_score}) within the ASE module. The weights of the edge between the item and its associated attributes are set to $1$. This approach allows for a nuanced representation of user preferences and their evolution over the course of the conversation.

We employ Graph Convolutional Network (GCN) ~\cite{kipf2016semi} to enhance the representations of nodes, denoted as $\mathcal{E}_{\text{node}}$. The primary objective is to capture evolving interrelationships within the current conversation context, $\mathcal{G}_u^{(t)}$. This process is mathematically represented as:
\begin{equation}
\label{eq:graph_i_conv}
    \mathbf{e}_{n}^{(l+1)}=\operatorname{ReLU}\left(\sum_{j \in \mathcal{N}_{n}^{(t)}} \frac{\mathbf{W}_{c}^{(l+1)} \mathbf{e}_{j}^{(l)}}{\sqrt{\sum_{i} \mathcal{E}_{n, i}^{(t)} \sum_{i} \mathcal{E}_{j, i}^{(t)}}}+\mathbf{e}_{n}^{(l)}\right),
\end{equation}
where $\mathbf{e}_{n}^{(l)}$ is the embedding for node $n$ at the $l^{th}$ GCN layer, 
$\mathcal{N}_(n)$ denotes the neighbors of node $n$ in $\mathcal{G}_u^{(t)}$, and $\mathbf{W}_{c}^{(l+1)} \in \mathbb{R}^{d\times d}$ represents the trainable weight matrix.

To encode the node representations corresponding to the user's clicking history, denoted as $\mathcal{P}_{\text {click}}$, we employ a Transformer  ~\cite{vaswani2017attention} architecture. This model allows us to effectively capture sequential patterns within the conversation history. Specifically, we leverage the Transformer to process and model this sequence. Finally, we derive the conversation state, $\mathbf{s}_{\text{conv}}^{(t)}$, by applying mean pooling to the node embeddings generated by the Transformer output. This process is illustrated as follows:
\begin{equation}
\label{eq:graph_s_conv}
    \mathbf{s}_{\text{conv}}^{(t)}= \text{MeanPool}(\text{Transformer}(\mathcal{E}_{\mathcal{P}_{\text {click}}})),
\end{equation}

Here, $\mathcal{E}_{\mathcal{P}{\text {click}}}$ denotes the sequence of GCN-encoded clicked attribute representations. The conversation state $\mathbf{s}_{\text{conv}}^{(t)}$ encapsulates relevant patterns from both the conversation context graph and user's historical clicks.

The graph-based conversation modeling module is designed to capture the user’s dynamic preferences and the system’s actions in a unified way. It provides a rich and flexible representation of the conversation state, which can be used to guide the policy learning module.

\subsubsection{Vague Preference Policy Learning} 
\label{method: unified_policy_learning}

We employ a Deep Q-Network (DQN) algorithm to address the challenge of making conversational decisions that consider users’ vague or dynamic preferences in CRS. The DQN algorithm can learn from the preference distributions estimated by ASE, which capture the user’s current and historical preferences over items and attributes. By optimizing the Q-value function, the DQN algorithm can predict the optimal action for each conversation state that maximizes the long-term conversational rewards.

The Q-value function $Q\left(s_{t}, a_{t}\right)$ of a policy $\pi$ is defined to measure the expectation of the accumulated rewards based on the state $s$ and the action $a$. We optimize the Q-function $Q^\ast \left(s_{t}, a_{t}\right)$:
\begin{equation}
\label{eq: Q-target}
    Q^*(s_t, a_t) = \max_{\pi} \mathbb{E}[R_{t+1} + \gamma \max_{a} Q^{\pi}(s_{t+1}, a) | \mathbf{s}_t, a_t]
\end{equation}
where $\pi$ is the policy, $R_{t+1}$ is the reward at turn $t+1$, $\gamma$ is the discount factor, and $Q^{\pi}(s_{t+1}, a)$ is the estimated action-value function for the next state and action.

However, the action space in CRS is usually large, as it includes all candidate attributes and items. Such a large action space can negatively impact the sampling efficiency and the convergence speed of the DQN algorithm. To enhance the sampling efficiency, we employ a preference-guided action pruning strategy. Specifically, we select the top-N items $\mathcal{V}_{\text{top}}^{(t)}$ and attributes $\mathcal{P}_{\text{top}}^{(t)}$ with the highest preference scores from ASE to construct a pruned action space. These preference scores reflect the user’s vague and dynamic preferences, and thus can help the agent focus on the most promising actions. However, setting a small value of $N$ may also limit the diversity and exploration of the agent. Therefore, we adopt the action space size configuration from previous work ~\cite{deng2021unified, zhang2022multiple}, setting it as $N=10$. The pruning action space is defined as:
\begin{equation}
\label{eq:a_sub}
    \mathcal{A}_{\text{action}}^{(t)} = \mathcal{V}_{\text{top-N}}^{(t)} + \mathcal{P}_{\text{top-N}}^{(t)} 
\end{equation}

For policy learning, the conversation state $\mathbf{s}_{\text{conv}}^{(t)}$ captures the user's dynamic conversation state, which is represented by a graph-based model as described in Section ~\ref{method: cur_conv_state}. The pruning action space $\mathcal{A}_{\text{action}}^{(t)}$ is determined by employing a preference-guided action pruning strategy, which partially estimate the user's vague preference distribution. The reward $R$ follows the previous MCR setting ~\cite{lei2020interactive}, and the detailed settings will be described in the Section ~\ref{Exp:traing_detail}.

\subsection{DQN Training}
Deep Q-Network (DQN) training is an essential process in enhancing the performance of reinforcement learning agents. To apply DQN to the Dynamism-aware Policy Learning (DPL) for VPPL, We adopt the same Dueling DQN and prioritized experience replay to optimize the Q-function $Q^\ast \left(s_{t}, a_{t}\right)$:

We initiate the DQN training by initializing a replay buffer $\mathcal{D}$ to store the transition tuples $(s_t, a_t, r_t, s_{t+1}, \mathcal{A}_{t+1})$. 
The DQN leverages a function approximator, typically a neural network, to estimate the Q-values, denoted as $Q(s,a;\theta)$ where $\theta$ are the parameters of the network.

During training, we sample a mini-batch of experiences from $\mathcal{D}$ and compute the loss as per the temporal-difference (TD) error:
\begin{equation}
\mathcal{L}(\theta) = \mathbb{E}_{(s,a,r,s')\sim\mathcal{D}}\left[ \left( r + \gamma \max_{a'} Q(s',a';\theta^-) - Q(s,a;\theta) \right)^2 \right],
\end{equation}
where $\theta^-$ are the parameters of a target network that are periodically updated with the weights of the online network, and $\gamma$ is the discount factor.

To address the overestimation of Q-values, we employ Double Q-learning, where the selection of the action in the target Q-value computation uses the online network, while the evaluation uses the target network:
\begin{equation}
y_t = r_t + \gamma Q\left(s_{t+1}, \argmax_{a'} Q(s_{t+1},a';\theta); \theta^-\right).
\end{equation}

The parameters of the target network $\theta^-$ are updated softly towards the parameters of the online network after every episode of interaction to stabilize training:
\begin{equation}
\theta^- \leftarrow \tau\theta + (1 - \tau)\theta^-,
\end{equation}
with $\tau$ being the update frequency.

Furthermore, to prioritize important transitions, we integrate a prioritized experience replay mechanism, sampling transitions with a probability proportional to their TD error, thus focusing the training on transitions from which there is more to learn.

The training proceeds iteratively by updating the Q-network parameters to minimize the TD error, thus enhancing the policy's capability to make informed decisions in the dynamic and complex setting of conversational recommendation systems.

\section{Experiments}

In this section, we evaluate the proposed method in VPMCR. We use the following research questions (RQs) to guide our experiment.
\begin{itemize}[leftmargin=1mm]
    \item \textbf{RQ1.} How does our VPPL method perform in comparison to state-of-the-art CRS methods in the VPMCR scenario?
    \item \textbf{RQ2.} How do the key components contribute to the overall performance of our VPPL method?
    \item \textbf{RQ3.} How well can the VPPL framework adapt to different CRS scenarios? Can it achieve superior performance in the MIMCR scenario as well?
     \item \textbf{RQ4.} How effective is our proposed method in modeling vague preferences under different vague preference initialization strategies?
    \item \textbf{RQ5.} How do the hyperparameters of our method affect its performance?
    \item \textbf{RQ6.} Can our method successfully make recommendations from users’ vague preference space?
\end{itemize}

\subsection{Dataset Description}
\begin{table}[t]
  \centering
  \caption{Statistics of datasets.}
  \resizebox{0.7\columnwidth}{!}{
    \begin{tabular}{lrrrr}
    \toprule
    \textbf{Dataset} & \multicolumn{1}{l}{\textbf{Yelp}} & \multicolumn{1}{l}{\textbf{LastFM}} & \multicolumn{1}{l}{\textbf{Amazon-Book}} & \multicolumn{1}{l}{\textbf{MovieLens}} \\
    \midrule
    \#Users &            27,675  &           1,801  &        30,291  &         20,892  \\
    \#Items &            70,311  &           7,432  &        17,739  &         16,482  \\
    \#Interactions &       1,368,609  &         76,693  &       478,099  &       454,011  \\
    \#Attributes &                590  &           8,438  &             988  &           1,498  \\
    \#Attribute-types &                  29  &               34  &               40  &               24  \\
    \midrule
    \#Entities &            98,576  &         17,671  &        49,018  &         38,872  \\
    \#Relations &                    3  &                 4  &                 2  &                 2  \\
    \#Triplets &       2,533,827  &       228,217  &       565,068  &       380,016  \\
    \bottomrule
    \end{tabular}%
    }
  \label{tab:data}%
\end{table}%

We introduce four datasets, whose statistics are shown in table ~\ref{tab:data}.
\begin{itemize}[leftmargin=1mm]
\item \textbf{Yelp and LastFM ~\cite{lei20estimation}:} 
Yelp\footnote{https://www.yelp.com/dataset/} and LastFM\footnote{https://grouplens.org/datasets/hetrec-2011/} datasets are used for business and music artist recommendations, respectively. 
We follow the multiple attribute question settings, retaining the original attribute instances in LastFM and Yelp, and extracting the attribute types they depend on. In Yelp, we utilize the 2-layer taxonomy designed by ~\cite{lei20estimation}, resulting in 29 categories in the first layer as attribute types and 590 attributes in the second layer as attribute instances. For LastFM, we follow ~\cite{zhang2022multiple}, retaining the original 8,438 attributes as attribute instances and employing clustering to obtain 34 attribute types.
\item \textbf{Amazon-Book ~\cite{wang2019kgat}:} Amazon Book\footnote{http://jmcauley.ucsd.edu/data/amazon.} is a widely used product recommendation dataset.  We retain users and items with at least 10 interaction records and consider entities (e.g., science fiction) and relations (e.g., genre) in the knowledge graph as attribute instances and attribute types, respectively.
\item \textbf{MovieLens:}
Movielens is a movie rating dataset. We adopt Movie\-Lens-20M\footnote{https://grouplens.org/datasets/movielens/} dataset, following ~\cite{zhang2022multiple}, and retain interactions with ratings greater than 3. We select entities and relations in the knowledge graph (KG) as attribute instances and attribute types, respectively.
\end{itemize}

\subsection{Experimental Setup}
\subsubsection{User Simulator in VPMCR}
\label{Exp:user_sim}

Conversational recommendation systems (CRSs) are interactive and require training and evaluation through user interactions. However, obtaining data directly from users in a research lab is impractical, so employing a user simulator is a common practice~\cite{chandramohan2011user}. The user simulator simulates users' interaction records in the training and test sets.

In the VPMCR scenario, we adopt a user simulation strategy similar to that in MIMCR~\cite{zhang2022multiple}, considering the reasonableness of the multi-interest setting. For a given observed user-items interaction pair $(u, \mathcal{V}_{u})$, we simulate a conversation session. Each item $v$ in $\mathcal{V}_{u}$ is treated as a ground-truth target item, and the union of attribute types and attributes associated with each item are considered as the user's ground-truth intent space $\mathcal{C}_{u}$ and ground-truth attribute space $\mathcal{P}$, respectively. The conversation session is initialized when the user specifies a common attribute $p_{0}$ to all $\mathcal{V}_{u}$, and the user's clear preference space $\mathcal{C}_{CI}$ and user's vague preference space $\mathcal{C}_{VI}$ are randomly initialized from the ground-truth intent space $\mathcal{C}_{u}$.

During the interaction, we use the ground-truth attribute space $\mathcal{P}$ as a criterion for the user simulator's acceptance or rejection. The detailed interaction process follows the ``system asks or recommends and user responds'' rules outlined in Section \ref{sec:definition}.

\subsubsection{Action Inference}
\label{method: action_inference}

The action inference involves either recommending items or asking an attribute-related question. 

(1) \textbf{Recommendation}: If an item $v$ in the action space has the highest Q-value, the CRS make a recommendation, resulting in a new action space $\mathcal{A}^{(t)} = \mathcal{V}_{top}^{(t)}$.

(2) \textbf{Questioning}: If an attribute $p$ in the action space has the highest Q-value, the CRS asks a question. In a multiple-choice setting, a two-level decision process is employed: first selecting an attribute type, then presenting several attributes within that type. A sum-based strategy ~\cite{zhang2022multiple} is used to determine the attribute type for questioning. Specifically, Q-values of all attributes within the attribute action space $\mathcal{P}_{top}^{(t)}$ are summed and allocated to their respective attribute types. The attribute type with the highest total value is selected for questioning, and the top $K$ attributes with the highest Q-values within that type are presented to the user.

\subsubsection{Baselines}
\label{Exp:baselines}
We use the following baselines. For fairness, all baselines are compared in the VPMCR scenario.
\begin{itemize}[leftmargin=1mm]
    \item \textbf{Max Entropy}. It selects the attribute with the maximum information entropy and inversely relates the probability of making a recommendation to the length of candidate items. While efficient, it lacks depth in capturing user-specific preferences, as it relies mainly on entropy metrics rather than on actual user feedback, which may reduce recommendation quality, particularly when the candidate pool is smaller.
    \item \textbf{CRM}~\cite{Sun:2018:CRS:3209978.3210002}. It employs a belief tracker to record user preferences as conversation state representation vectors, that are then applied to reinforcement learning decision module and factorization machine (FM) recommendation modules, respectively.  However, CRM’s reliance on state vectors can lead to shallow preference modeling, limiting its capacity to capture deeper nuances in user feedback and adapt effectively to dynamic user preferences.
    \item \textbf{EAR}~\cite{lei20estimation}. This method adopts the three-stage solution framework to enhance the interaction between the conversation component and the recommendation component.  EAR using synthetic user behavior (modeled via Max Entropy) for factorization machine (FM) offline training. This reliance on entropy-modeled user behavior may lead to oversimplified FM modeling, which limits the framework's ability to capture the full complexity of user feedback, especially by only treating clicked attributes as positive signals, overlooking the significance of unclicked attributes.
    \item \textbf{SCPR}~\cite{lei2020interactive}. SCPR leverages graph-based path reasoning to prune useless candidate attributes. It separates attribute selection from reinforcement learning, which is only used for determining when to ask and recommend. Despite optimizing decision efficiency, SCPR shares EAR’s offline FM recommendation model, facing similar drawbacks in  preference modeling due to its limited consideration of ambiguous feedback signals.
    \item \textbf{UNICORN}~\cite{deng2021unified}. A state-of-the-art method for the MCR scenario that proposes a unified policy learning framework using dynamic graphs to model conversation states and employs a preference-based scoring to reduce reinforcement learning action space. However, it interprets user feedback in a binary manner, overlooking the subtleties in user preferences that fall between clear acceptances and rejections. UNICORN treats historical user clicks as positive signals and non-clicks as negative signals, failing to account for the time-sensitive shifts in user feedback and ignoring the influence of recent feedback in shaping ongoing preferences.
    \item \textbf{MCMIPL}~\cite{zhang2022multiple}. It considers the user's multi-interest space and extends the MCR scenario to a more realistic MIMCR scenario. This method also follows the graph-based unified reinforcement learning framework and employs the multi-interest encoder to learn the conversation state. However, like UNICORN, MCMIPL relies on preference-based scoring for decision space reduction, also interpreting non-clicked attributes as negative signals. This limitation hinders MCMIPL’s ability to adapt to the dynamic and sometimes ambiguous nature of user preferences, which our approach explicitly addresses.
\end{itemize}

\subsubsection{Training Details}
\label{Exp:traing_detail}

In our experiments, each dataset was divided into training, validation, and testing sets with a 7:1.5:1.5 ratio. For the user simulator, the maximum conversation turn, \( T \), was set to 15, and each user was assigned two target item sets, denoted as \( \mathcal{V}_u \). User preference spaces, both vague and clear, were initialized using uniform sampling.

The Ambiguity-aware Soft Estimation (ASE) module settings included intensity coefficients \( \lambda_1 = 0.1 \) and \( \lambda_2 = 0.01 \), with a decay discount factor of 0.1. In the Dynamism-aware Policy Learning (DPL) module, we employed random sampling for candidate item selection in constructing the dynamic graph when the available candidates exceeded 5000. The conversation modeling architecture comprised two Graph Neural Network (GNN) layers and one Transformer layer, with embedding and hidden sizes set at 64 and 100, respectively. For action pruning in Reinforcement Learning (RL), the item and attribute space sizes were set to 10 (\( N = 10 \)), and the number of attributes displayed to the user was limited to 2 (\( K = 2 \)). Graph node embeddings were pre-trained using TransE. ~\cite{bordes2013translating}. 

Our Deep Q-Network (DQN) was trained over 10,000 episodes, with reward settings consistent with benchmarks: \( r_{\text{rec-suc}} = 1 \), \( r_{\text{rec-fail}} = -0.01 \), \( r_{\text{ask-suc}} = -0.1 \), \( r_{\text{ask-fail}} = -0.1 \), and \( r_{\text{quit}} = -0.3 \). The experience replay buffer size was set to 50,000, with a mini-batch size of 128. Optimization was performed using the Adam algorithm, with a learning rate of \( 1e-4 \) and L2 regularization of \( 1e-6 \).

In the RL setup, rewards were structured to encourage accurate recommendations and efficient conversation. Positive rewards were given for successful recommendations (\( r_{\text{rec-suc}} \)), while penalties were imposed for unsuccessful recommendations (\( r_{\text{rec-fail}} \)), overabundance questions (\( r_{\text{ask-suc}} \) and irrelevant questions \( r_{\text{ask-fail}} \)), and system termination (\( r_{\text{quit}} \)).

\subsubsection{Evaluation Metrics}
\label{Exp:evaluation_metric}
We evaluate performance using Success Rate (SR@$T$) and Average Turn
(AT). SR@$T$ measures the percentage of successful recommendations within $T$ turns; higher is better. AT measures the average conversation length; lower indicates greater efficiency.

We use hierarchical normalized Discounted Cumulative Gain (hDCG@($T, K$)) to evaluate the ranking of the top-$K$ recommendations within $T$ turns. hDCG assigns higher scores to recommendations that are more relevant to the user. A higher hDCG@($T, K$) indicates a better ranking performance.
\begin{table*}[t]
  \large
  \centering
  \renewcommand{\arraystretch}{2} 
  \caption{Performance comparison of different models in in the Vague Preference Multi-round Conversational Recommendation (VPMCR) scenario. hDCG stands for hDCG@(15, 10). \textbf{Bold number} represents the improvement of VPPL over existing models is statistically significant ($p<0.01$). }
  \resizebox{\columnwidth}{!}{
    \begin{tabular}{cccccccccccccccc}
    \toprule
    \multirow{2}[4]{*}{\textbf{Models}} & \multicolumn{3}{c}{\textbf{Yelp}} &       & \multicolumn{3}{c}{\textbf{LastFM}} &       & \multicolumn{3}{c}{\textbf{Amazon-Book}} &       & \multicolumn{3}{c}{\textbf{MovieLens}} \\
\cmidrule{2-4}\cmidrule{6-8} \cmidrule{10-12}\cmidrule{14-16}         & \textbf{SR@15} & \textbf{AT} & \textbf{hDCG} &       & \textbf{SR@15} & \textbf{AT} & \textbf{hDCG} &       & \textbf{SR@15} & \textbf{AT} & \textbf{hDCG} &       & \textbf{SR@15} & \textbf{AT} & \textbf{hDCG} \\
    \midrule
    Max Entropy & 0.062  & 14.44  & 0.030  &       & 0.376  & 11.25  & 0.189  &       & 0.180  & 12.91  & 0.107  &       & 0.448  & 9.93  & 0.315  \\
    CRM   & 0.212  & 13.27  & 0.070  &       & 0.372  & 12.26  & 0.126  &       & 0.296  & 12.34  & 0.109  &       & 0.780  & 5.96  & 0.341  \\
    EAR   & 0.232  & 13.05  & 0.080  &       & 0.414  & 11.61  & 0.146  &       & 0.324  & 12.14  & 0.119  &       & 0.792  & 5.50  & 0.361  \\
    SCPR  & 0.322  & 12.34  & 0.115  &       & 0.596  & 10.18  & 0.206  &       & 0.374  & 11.62  & 0.139  &       & 0.806  & 4.90  & 0.387  \\
    UNICORN & 0.314  & 12.11  & 0.140  &       & 0.632  & 9.17  & 0.280  &       & 0.396  & 11.05  & 0.193  &       & 0.810  & 4.81  & 0.548  \\
    MCMIPL & 0.322  & 12.16  & 0.136  &       & 0.634  & 9.52  & 0.267  &       & 0.412  & 10.90  & 0.205  &       & 0.820  & 4.39  & 0.579  \\
\midrule
\textbf{VPPL} & \textbf{0.398 } & \textbf{11.26 } & \textbf{0.175 } &       & \textbf{0.686 } & \textbf{8.58 } & \textbf{0.306 } &       & \textbf{0.424 } & \textbf{10.75 } & \textbf{0.206 } &       & \textbf{1.000 } & \textbf{1.60 } & \textbf{0.689 } \\
\bottomrule
    \end{tabular}%
    }
  \label{tab:performance}%
\end{table*}%

\begin{figure}[htbp]
  \centering
  \begin{minipage}[t]{0.47\linewidth}
    \centering
    \includegraphics[width=\linewidth]{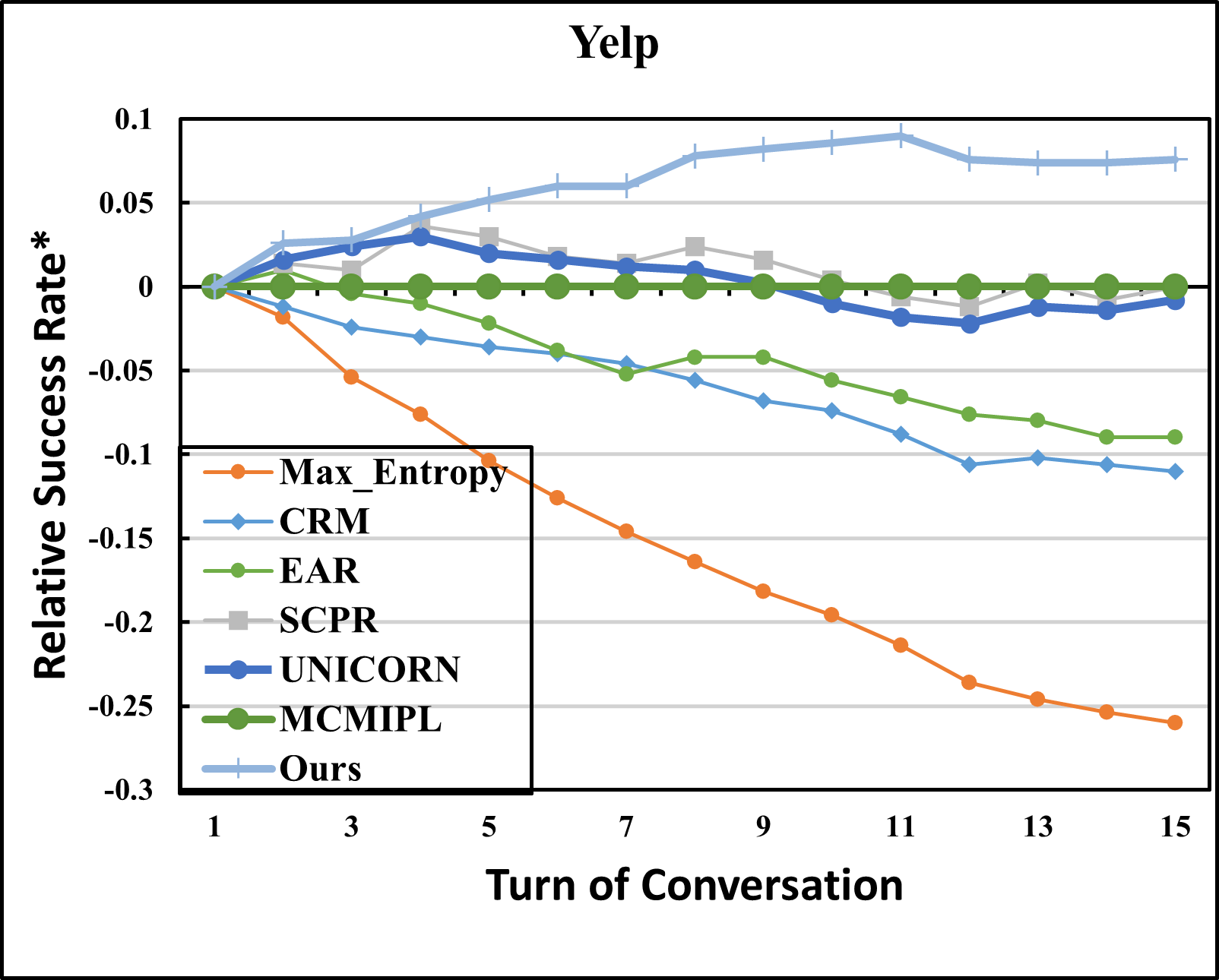}
    \label{fig:yelp}
  \end{minipage}%
  \hfill%
  \begin{minipage}[t]{0.47\linewidth}
    \centering
    \includegraphics[width=\linewidth]{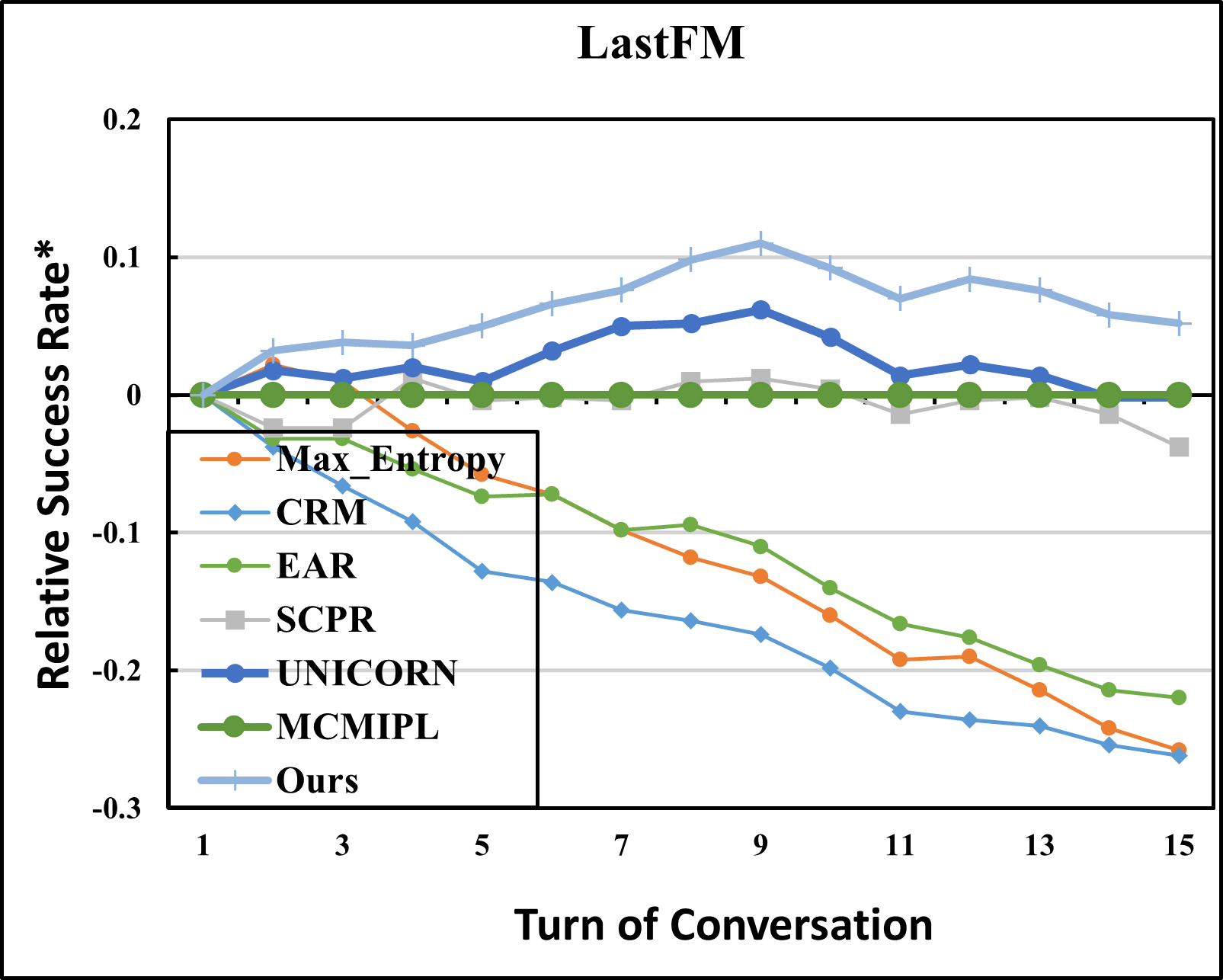}
    \label{fig:lastfm}
  \end{minipage}

  \vspace{-0.2cm} 
  
  \begin{minipage}[t]{0.47\linewidth}
    \centering
    \includegraphics[width=\linewidth]{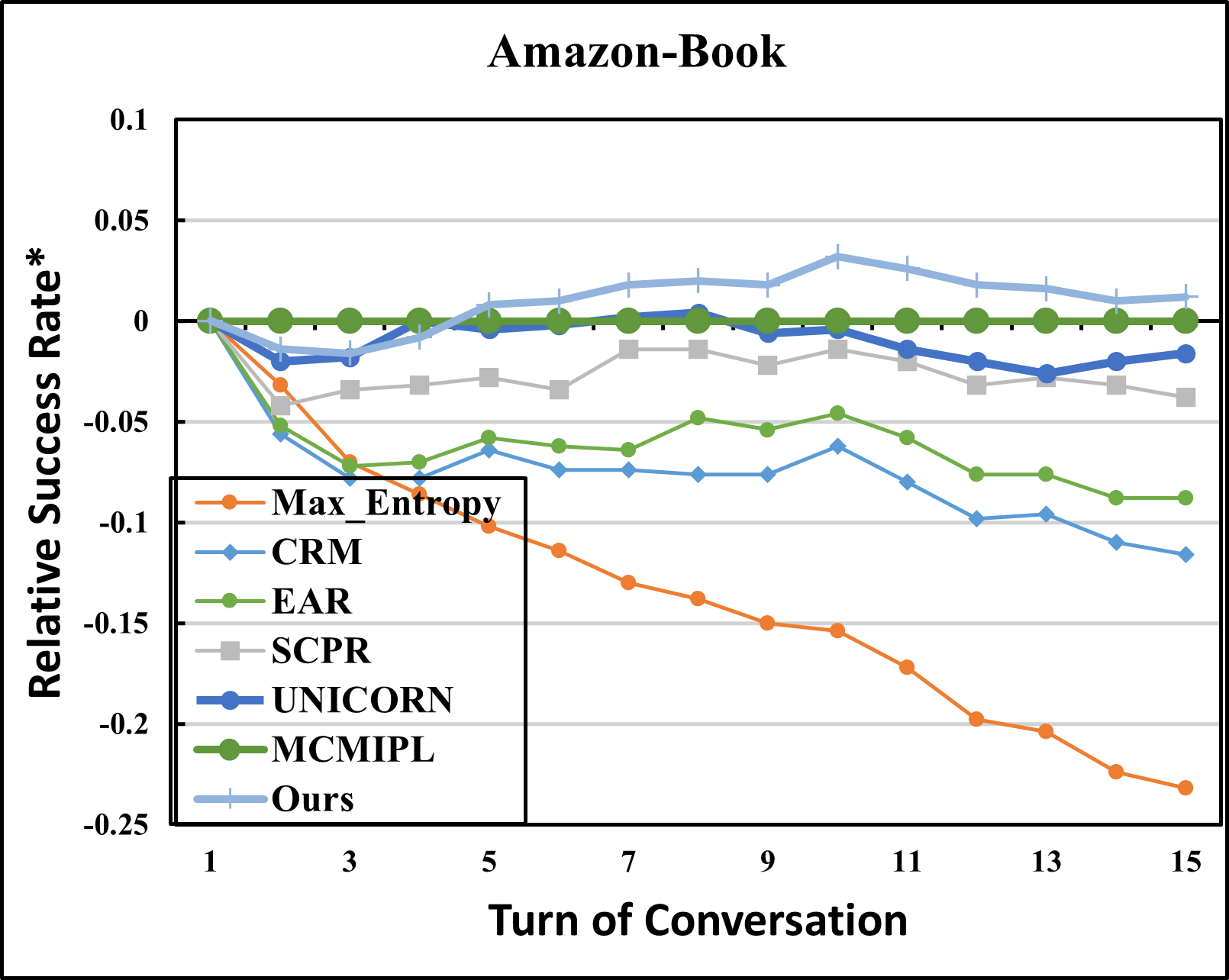}
    \label{fig:book}
  \end{minipage}%
  \hfill%
  \begin{minipage}[t]{0.47\linewidth}
    \centering
    \includegraphics[width=\linewidth]{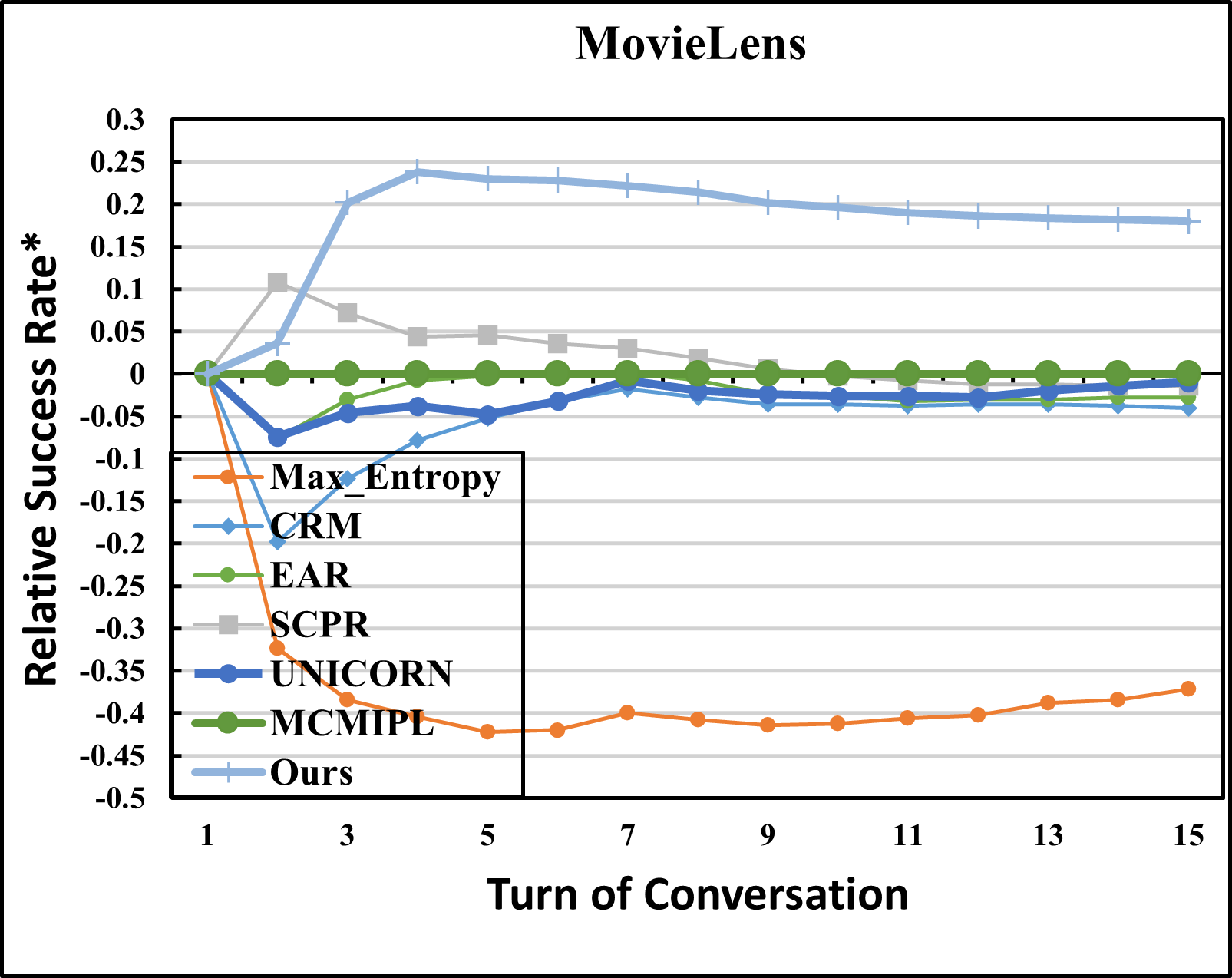}
    \label{fig:movielens}
  \end{minipage}
  \caption{SR* of compared methods at different turns on four datasets (RQ1)}
  \label{fig:fourplots_RQ1}
\end{figure}

\subsection{Performance comparison of VPPL with existing models (RQ1)}
In addressing RQ1, we delve into the comparative performance of the Vague Preference Policy Learning (VPPL) model against established baseline models in the Vague Preference Multi-round Conversational Recommendation (VPMCR) scenario.

Table \ref{tab:performance} reports the Success Rate@$15$(SR@$15$), Average Turn (AT) and hierarchical Discounted Cumulative Gain (hDCG@($15, 10$)) for VPPL and baseline models. The statistics in Table \ref{tab:performance} show that VPPL outperforms all baselines with a higher success rate and a less average turn, especially on Yelp and Movielens. This superiority is attributed to VPPL's nuanced ability to model vague user preferences and accommodate dynamic preference shifts, a key limitation in existing models. Specifically, VPPL's Ambiguity-aware Soft Estimation (ASE) module is able to capture both directed and inferred preferences, whereas traditional baselines rely heavily on binary user responses, which may lead to over-filtering or misinterpreting user preferences. This ability to handle the inherent vagueness and variability in user feedback sets VPPL apart from other models, which often struggle in complex user interaction scenarios where preferences are not clear-cut.

Figure \ref{fig:fourplots_RQ1} shows the relative success rate (SR*) of each model at every turn compared to the MCMIPL baseline (represented by the dark green line at $y=0$). Observing the variation trend of curves in Figure \ref{fig:fourplots_RQ1}, we have the following findings:

\begin{itemize}[leftmargin=1mm]
    \item VPPL almost consistently and substantially surpassed all baselines over the entire conversation session across datasets. Specifically, VPPL achieved a high success rate in the first a few turns on MovieLens, demonstrates its proficiency in swiftly converging towards user-preferred items, capturing vague preferences more effectively than baselines that use fixed decision boundaries.
    \item As the conversation continues, the performance gaps between VPPL and other baselines expand, especially in the case of Max Entropy. The Max Entropy model lacks an adaptive decision-making mechanism that leverages nuanced feedback from users, leading to an overly long and inefficient dialogue process. In contrast, VPPL uses reinforcement learning with dynamic graphs to predict optimal actions, allowing for adaptive exploration of user preferences. This adaptiveness enables VPPL to expedite the conversation with more targeted questioning and reduce average turns. 
    \item Baselines like CRM and EAR rely on state representations that may inadequately capture user preference evolution throughout the conversation. For example, the belief tracker in CRM operates based on simplistic binary feedback, thus failing to effectively represent complex and ambiguous preference shifts. In contrast, VPPL benefits from its graph representation of conversation state and its preference distribution, allowing a more granular understanding of user behavior, particularly where preferences are dynamic or vaguely expressed.
    \item  Reinforcement learning-based methods like CRM and EAR lag behind more advanced models, as they directly apply RL to a large decision space without effectively representing the conversation state, hindering optimal policy learning. The SCPR, UNICORN, and MCMIPL models, which leverage graph-based representations for more efficient decision-making, still underperform compared to VPPL due to their reliance on binary preference scoring. VPPL, by treating non-clicks as potentially meaningful data points rather than outright rejections, manages to preserve diversity in the candidate set, thereby better adapting to dynamic user needs.

\end{itemize}
\vspace{-5pt}
\subsection{Evaluating Key Design in VPPL (RQ2)}
\begin{table*}[ht]
\large
  \centering
  \renewcommand{\arraystretch}{2} 
  \caption{Ablation study of VPPL components in the Vague Preference Multi-round Conversational Recommendation (VPMCR) scenario.}
  \resizebox{\textwidth}{!}{
    \begin{tabular}{ccccrcccrcccrccc}
    \toprule
    \multirow{2}[4]{*}{} & \multicolumn{3}{c}{\textbf{Yelp}} &       & \multicolumn{3}{c}{\textbf{LastFM}} &       & \multicolumn{3}{c}{\textbf{Amazon-Book}} &       & \multicolumn{3}{c}{\textbf{MovieLens}} \\
\cmidrule{2-4}\cmidrule{6-8}  \cmidrule{10-12}  \cmidrule{14-16}        & \textbf{SR@15} & \textbf{AT} & \textbf{hDCG} &       & \textbf{SR@15} & \textbf{AT} & \textbf{hDCG} &       & \textbf{SR@15} & \textbf{AT} & \textbf{hDCG} &       & \textbf{SR@15} & \textbf{AT} & \textbf{hDCG} \\
    \midrule
    \textbf{VPPL - (VPMCR)} & \textbf{0.398} & \textbf{11.26} & \textbf{0.175} &       & \textbf{0.686} & \textbf{8.58} & \textbf{0.306} &       & \textbf{0.424} & \textbf{10.75} & \textbf{0.206} &       & \textbf{1.000 } & 1.60   & 0.689 \\
    \midrule
    \multicolumn{1}{l}{(a) - w/o ASE Item.Score} & 0.328  & 12.04  & 0.144  &       & 0.618  & 9.35  & 0.271  &       & 0.386  & 11.17  & 0.189  &       & 0.852  & 3.84  & 0.593  \\
    \multicolumn{1}{l}{(b) - w/o ASE Attr.Score} & 0.354  & 11.88  & 0.149  &       & 0.614  & 9.44  & 0.267  &       & 0.412  & 10.91  & 0.199  &       & 1.000  & 1.75  & 0.663  \\
    \midrule
    \multicolumn{1}{l}{(c) - w/o Personalized Preference} & 0.142  & 13.84  & 0.060  &       & 0.444  & 10.79  & 0.211  &       & 0.284  & 12.10  & 0.142  &       & 0.858  & 5.22  & 0.492  \\
    \multicolumn{1}{l}{(d) - w/o Average Preference} & 0.368  & 11.38  & 0.169  &       & 0.630  & 9.24  & 0.269  &       & 0.416  & 10.84  & 0.199  &       & 1.000  & 1.77  & 0.668  \\
    \multicolumn{1}{l}{(e) - w/o Decaying Preference} & 0.382  & 11.56  & 0.163  &       & 0.628  & 9.15  & 0.280  &       & 0.410  & 11.05  & 0.190  &       & 1.000  & \textbf{1.49 } & \textbf{0.708 } \\
    \bottomrule
    \end{tabular}%
    }
  \label{tab:ab_study}%
\end{table*}%

\subsubsection{Key Components of VPPL}
We assess the efficacy of the Ambiguity-aware Soft Estimation (ASE) within our VPPL framework, particularly in its role for navigating conversations and accommodating shifts in user preferences within various VPMCR contexts. We separately remove the ASE module for items and attributes (Section ~\ref{method: ASE}) and replace them with a preference-based scoring strategy ~\cite{deng2021unified, zhang2022multiple}, which models user preferences using historical click or non-click attributes as mixed signals.

Results from the ablation study, presented in Table \ref{tab:ab_study} (rows a-b), clearly demonstrate a marked decline in performance across all datasets when the ASE module is excluded. This decline clearly indicates that the ASE's ability to distinguish between directed and inferred preferences is crucial for modeling user vagueness effectively. Without ASE, VPPL becomes similar to existing models like UNICORN and MCMIPL that fail to capture implicit preferences from non-clicked options, leading to suboptimal filtering and a lower success rate.

Furthermore, our analysis reveals a notable aspect of ASE: its pronounced effectiveness in discerning user preferences towards items over attributes. This distinction can be attributed to the more direct and explicit nature of click behaviors, which act as a clearer indicator of user preferences for items.

\subsubsection{Key Components of ASE}

In assessing the integral components of the Ambiguity-aware Soft Estimation (ASE) module, we executed a series of ablation experiments, the results of which are detailed in Table \ref{tab:ab_study}, rows (c-e). These experiments were meticulously designed to discern the individual contributions of various ASE components to the overall framework efficacy.

Row (c) of Table \ref{tab:ab_study} shows the paramount importance of personalized user modeling within ASE. The absence of this feature results in the model's inability to capture individualized user preferences, thereby significantly hampering overall performance. This finding underscores the necessity of personalized preference modeling in crafting effective conversational recommendation systems.

In Row (d), the ablation of the average preference, as defined in Equation \ref{eq:avg_score}, manifests in performance decline across all datasets. Notably, the LastFM dataset exhibited the most pronounced degradation, potentially attributable to its numerous attributes and the substantial role of non-displayed attribute information in shaping user preference predictions.

Furthermore, Row (e) examines the effect of omitting the historical decay preference component of the time-aware preference decay. This exclusion led to a decline in performance across three datasets, with the notable exception of MovieLens. Intriguingly, on the MovieLens dataset, the ASE module, bereft of decaying historical information, demonstrated a robust ability to estimate current turn preferences, with recommendations proving effective within the initial few interaction rounds. This observation suggests that the integration of historical decay preference in scenarios characterized by brief interactive sequences might inadvertently dilute the precision of preference inference, particularly in the context of the MovieLens dataset.

Collectively, these results not only affirm the criticality of the ASE module within the VPPL framework but also provide insightful evidence of the framework's versatility and effectiveness in diverse conversational recommendation scenarios.

\begin{table*}[ht]
\large
  \centering
  \renewcommand{\arraystretch}{2} 
  \caption{Comparative analysis of VPPL and other baselines in the Multi-Interest Multi-round Conversational Recommendation (MIMCR) scenario.}
  \resizebox{\textwidth}{!}{
    \begin{tabular}{ccccrcccrcccrccc}
    \toprule
    \multirow{2}[4]{*}{} & \multicolumn{3}{c}{\textbf{Yelp}} &       & \multicolumn{3}{c}{\textbf{LastFM}} &       & \multicolumn{3}{c}{\textbf{Amazon-Book}} &       & \multicolumn{3}{c}{\textbf{MovieLens}} \\
\cmidrule{2-4}\cmidrule{6-8}  \cmidrule{10-12}  \cmidrule{14-16}        & \textbf{SR@15} & \textbf{AT} & \textbf{hDCG} &       & \textbf{SR@15} & \textbf{AT} & \textbf{hDCG} &       & \textbf{SR@15} & \textbf{AT} & \textbf{hDCG} &       & \textbf{SR@15} & \textbf{AT} & \textbf{hDCG} \\
    \midrule
    \textbf{VPPL} & \textbf{0.636} & \textbf{10.68} & \textbf{0.210} &       & 0.840  & 7.33  & \textbf{0.350} &       & \textbf{0.610} & \textbf{9.81} & \textbf{0.251} &       & \textbf{0.988} & \textbf{2.42} & \textbf{0.640} \\
    \midrule
    MCMIPL & 0.552  & 10.95  & 0.204  &       & \textbf{0.856 } & \textbf{7.21 } & 0.342  &       & 0.544  & 10.32  & 0.239  &       & 0.838  & 4.23  & 0.602  \\
    UNICORN & 0.454  & 11.01  & 0.188  &       & 0.832  & 7.42  & 0.350  &       & 0.530  & 10.23  & 0.231  &       & 0.832  & 4.35  & 0.567  \\
    SCPR & 0.452  & 12.52  & 0.136  &       & 0.688  & 10.27  & 0.220  &       & 0.450  & 11.10  & 0.167  &       & 0.834  & 4.80  & 0.392  \\
    \bottomrule
    \end{tabular}%
    }
  \label{tab:scenarios_cmp}%
\end{table*}%

\subsection{Performance Comparison in MIMCR Scenario (RQ3)}

In this study, we extend the evaluation of the Vague Preference Policy Learning (VPPL) framework to the Multi-Interest Multi-round Conversational Recommendation (MIMCR) scenario. MIMCR represents scenarios where users possess clear preferences, explicitly signaling acceptance or rejection of attributes pertaining to the desired item. This model capitalizes on user interactions involving multi-choice questions, utilizing user feedback as effective filters in the item selection process. 

Our comparative analysis, presented in Table \ref{tab:scenarios_cmp}, showcases the performance of VPPL against established baselines within the MIMCR scenario. Notably, VPPL demonstrates marked superiority in handling datasets from Yelp, Amazon-Book, and Movielens. While it exhibits a marginally lower performance in the SR and AT  metrics compared to the MCMIPL model on the LastFM dataset, VPPL outstrips all competitors in terms of hDCG, highlighting its robustness.

These findings corroborate the efficacy of VPPL in discerning user preferences in scenarios involving multi-choice questions. They underscore the model's versatility and its adeptness in accommodating both VPMCR and MIMCR scenarios. This adaptability enables VPPL to handle both well-defined user preferences as well as vague, evolving preferences, thereby achieving a commendable performance across various CRS environments.

\subsection{Vague Preference Initialization Strategies (RQ4)}

\begin{figure}[h]
\begin{minipage}[t]{0.49\linewidth}
    \includegraphics[width=\linewidth]{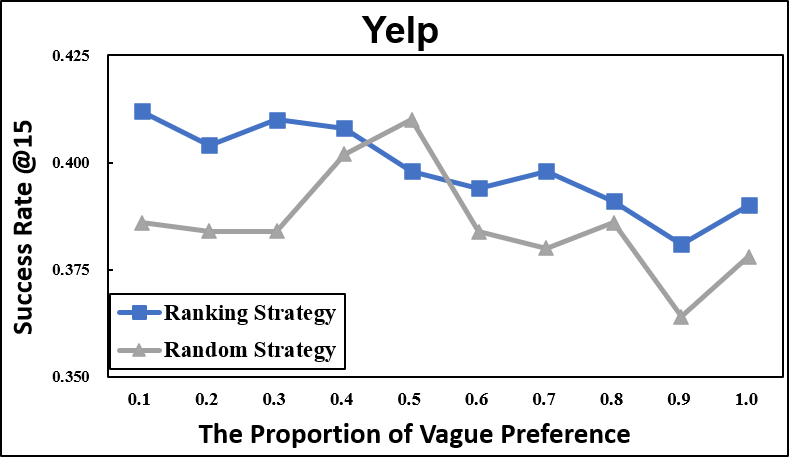}
    \label{f1}
\end{minipage}%
    \hfill%
\begin{minipage}[t]{0.49\linewidth}
    \includegraphics[width=\linewidth]{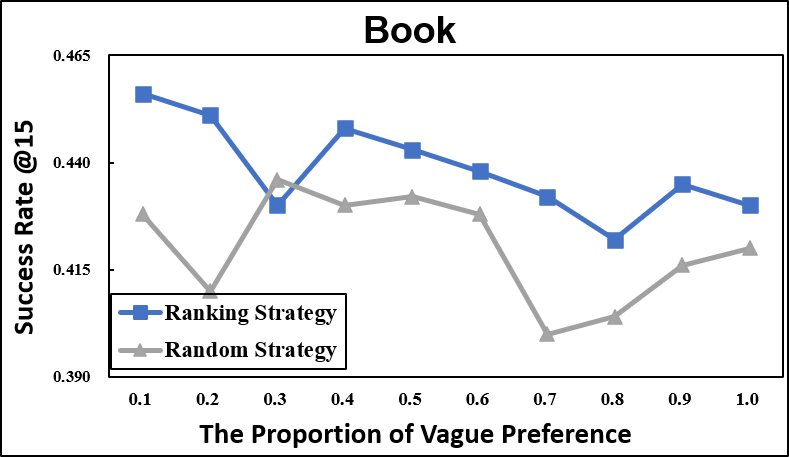} 
    \label{f2}
\end{minipage}
\caption{Performance (SR@15) of the Random and Ranking strategies for vague preference initialization under varying proportions of vague preferences on the Yelp (left) and Amazon-Book (right) datasets.
} 
\label{fig:vague_init}
\end{figure}

We conducted an additional experiment exploring the performance under different vague preference initialization strategies. In the previous experiments, users' vague preferences were randomly sampled from the preference space, without considering personalized user preference. However, in real-world scenarios, users' vague preferences may be influenced by their individual affinities toward certain attribute types.

To simulate this, we initialized vague preferences based on user affinity scores for attribute types, leveraging TransE \cite{moon-2019-opendialkg}, a graph embedding technique, to learn vector representations for users and attributes. User affinity scores for attributes were computed as the inner product of the user and attribute vectors, e.g., $w_{p\mbox{-}u} = e_{u}^{\top} e_p$, representing the user's interest level for that attribute. Attribute types with higher affinity scores were more likely to be clear preferences for the user, while those with lower scores were considered vague preferences. Consequently, we treated the attribute types with the lowest affinity scores as the user's vague preferences and explored the performance (SR@15) of two strategies: (1) Random strategy: Vague preferences were randomly sampled from the preference space, without considering user affinity. (2) Ranking Strategy: Vague preferences were constructed based on the user's affinity scores for attribute types, with the lowest-ranked attribute types considered vague preferences.

We evaluated the performance under varying proportions of vague preferences, ranging from 0.1 to 1.0 (i.e., all preferences are vague). We selected two representative datasets, Yelp and Amazon-Book, for this experiment. Figure\ref{fig:vague_init} depicts the performance (SR@15) of the two strategies under varying proportions of vague preferences.

The results show that the Ranking Strategy generally outperforms the Random Strategy, indicating that personalizing vague preferences based on user affinity towards attribute types improves the effectiveness of VPPL. As the proportion of vague preferences increases, the performance of the Ranking Strategy declines for both datasets, while the Random Strategy exhibits a non-monotonic behavior, with peak performance around a 0.4-0.5 proportion of vague preferences.

These findings highlight the importance of considering personalized user affinities when initializing vague preferences. However, as the proportion of vague preferences increases, more sophisticated methods may be required to accurately disentangle clear and vague preferences based on user behavior and affinity signals.

\subsection{Model Parameter Analysis (RQ5)}
In this section, we analyze the impact of the hyperparameters in the ASE module, which is the core component of our VPPL method in the VPMCR scenario. We follow the same experimental settings as in the previous section, and compare the success rate under different hyperparameter values. We omit the analysis of the hyperparameters in the policy learning module, as they have been extensively studied in the previous work on graph-based policy learning \cite{deng2021unified}.

\subsubsection{Hyperparameter Analysis in ASE}
We investigate two key hyperparameters in the ASE module:

\begin{table}[htbp]
  \centering
  \renewcommand{\arraystretch}{2} 
  \caption{The impact of the coefficient of information intensity w.r.t. SR@15.}
  \resizebox{\textwidth}{!}{
    \begin{tabular}{cc|ccccccccccccccc}
    \toprule
    \multicolumn{2}{c}{\textbf{Dataset}} & \multicolumn{3}{c}{\textbf{Yelp}} &       & \multicolumn{3}{c}{\textbf{LastFM}} &       & \multicolumn{3}{c}{\textbf{Amazon-Book}} &       & \multicolumn{3}{c}{\textbf{MovieLens}} \\
\cmidrule{3-5}\cmidrule{7-9}\cmidrule{11-13}\cmidrule{15-17}    \multicolumn{2}{c}{\textbf{$\lambda_2$}} & 0.01  & 0.1   & 1     &       & 0.01  & 0.1   & 1     &       & 0.01  & 0.1   & 1     &       & 0.01  & 0.1   & 1 \\
    \midrule
    \multirow{3}[2]{*}{\textbf{$\lambda_1$}} & 0.01  & \textbf{0.414 } & 0.408  & 0.328  &       & 0.592  & 0.646  & 0.592  &       & 0.424  & \textbf{0.430 } & 0.400  &       & \textbf{1.000 } & \textbf{1.000 } & \textbf{1.000 } \\
          & 0.1   & 0.398  & 0.410  & 0.344  &       & \textbf{0.686 } & 0.634  & 0.508  &       & 0.424  & 0.414  & 0.384  &       & \textbf{1.000 } & \textbf{1.000 } & 0.996  \\
          & 1     & 0.394  & 0.370  & 0.302  &       & 0.604  & 0.572  & 0.528  &       & 0.420  & 0.398  & 0.406  &       & \textbf{1.000 } & 0.996  & 0.992  \\
    \bottomrule
    \end{tabular}%
  }
  \label{tab:para_analysis}%
\end{table}%

The information intensity coefficients $\lambda_1$ and $\lambda_2$ determine the relative importance of directed preferences and inferred preferences in the preference extraction process. As shown in Table \ref{tab:para_analysis}, larger $\lambda_1$ and smaller $\lambda_2$ lead to higher success rates, implying that directed preferences ($\lambda_1$) are more informative and reliable than inferred preferences ($\lambda_2$) in the VPMCR scenario. However, when both $\lambda_1$ and $\lambda_2$ are large, the performance drops significantly, especially for sparser datasets like Yelp, indicating that the model is sensitive to the noise and ambiguity in the user feedback.

\begin{figure}[h]
\begin{minipage}[t]{0.49\linewidth}
    \includegraphics[width=\linewidth]{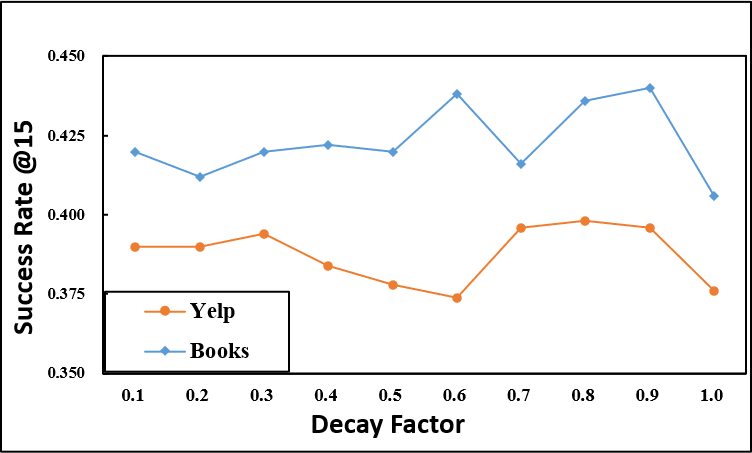}
    \label{f1}
\end{minipage}%
    \hfill%
\begin{minipage}[t]{0.49\linewidth}
    \includegraphics[width=\linewidth]{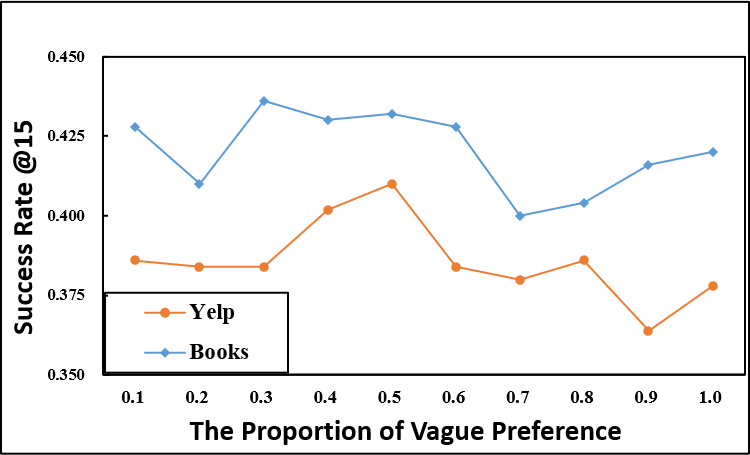} 
    \label{f2}
\end{minipage}
\caption{Comparative performance analysis of Success Rate with varying decay factor (left) and proportion of vague preference (right) hyperparameters.
} 
\label{fig:RQ3}
\end{figure}


The decay factor $\gamma$ adjusts the influence of historical preferences on the current preference estimation. Due to the performance variance across different datasets, we only present the results for Yelp and Amazon-Book, but note that LastFM and Movielens follow similar trends. As illustrated in Figure \ref{fig:RQ3}(left), a moderate decay factor (0.6-0.8) achieves the best performance, suggesting that a balance between recent and historical preferences is beneficial for capturing the user’s dynamic preference changes. On the other hand, extreme values (0.1 and 1.0) result in poor performance, indicating that either ignoring historical preferences or relying solely on recent preferences is detrimental for the VPMCR scenario.

\subsubsection{Proportion of Vague Preferences}
We varied the proportion of vague preferences in the user profiles from 0.1 to 1 and measured the performance to investigate the impact of the proportion of vague preferences.

As shown in Figure \ref{fig:RQ3}(right), the success rate was highest when the proportion of vague preferences was moderate (around 0.4 to 0.5). This indicates that our method can effectively leverage both vague and clear preferences to generate personalized recommendations. However, when the proportion of vague preferences was too high (over 0.7 to 0.8), the success rate dropped significantly. This suggests that our method has difficulty in inferring the user’s true intentions from too many vague preferences, resulting in less accurate recommendations.

\subsubsection{Impact of Pruning Action Space Size}
\begin{table}[htbp]
  \centering
  \caption{Impact of Action Space Size on Performance}
  \resizebox{0.6\textwidth}{!}{
    \begin{tabular}{cc|ccccccc}
    \toprule
    \multicolumn{2}{c}{\textbf{Dataset}} & \multicolumn{3}{c}{\textbf{Yelp}} &       & \multicolumn{3}{c}{\textbf{Amazon-Book}} \\
\cmidrule{3-5}\cmidrule{7-9}    \multicolumn{2}{c}{ $\left|\mathcal{V}_{\text{top-N}}\right|$} & 10    & 20    & 50    &       & 10    & 20    & 50 \\
    \midrule
    \multirow{3}[2]{*}{\textbf{$\left|\mathcal{P}_{\text{top-N}}\right|$}} & 1     & 0.380  & \textbf{0.388 } & 0.370  &       & 0.462  & \textbf{0.470 } & 0.446  \\
          & 10    & 0.362  & 0.362  & 0.340  &       & 0.416  & 0.448  & 0.401  \\
          & 20    & 0.320  & 0.335  & 0.292  &       & 0.392  & 0.388  & 0.360  \\
    \bottomrule
    \end{tabular}%
    }
  \label{tab:action_size}%
\end{table}%
In response to concerns regarding the configuration of our pruned action space in conversational recommendation systems, we conducted extensive experiments to assess the impact of varying the dimensions of the action space. Our experiments, focused on the Yelp and Amazon-Book datasets, demonstrate how different settings of \(N_{item}\) and \(N_{attr}\) influence system performance.

The results in Table \ref{tab:action_size} show that when both the item and attribute action spaces are relatively large (e.g., $N_{item}=50$, $N_{attr}=20$), the model performance tends to decrease. On the other hand, when the attribute action space is smaller ($N_{attr}=1$) and the item action space is moderate ($N_{item}=20$), the model achieves the best performance on both datasets.

This observation suggests that a larger action space may not necessarily be better for learning vague user intentions, as it could introduce more noise and complexity, potentially hindering the model's ability to accurately capture the nuances of user preferences. A more balanced and focused action space, with a moderate number of top-ranked items and a smaller number of top-ranked attributes, appears to be more effective in the VPMCR scenario, where the goal is to adapt to users' vague and dynamic preferences.

\subsubsection{Impact of Node Sampling on Efficiency and Effectiveness}
\begin{figure}[h]
\begin{minipage}[t]{0.49\linewidth}
    \includegraphics[width=\linewidth]{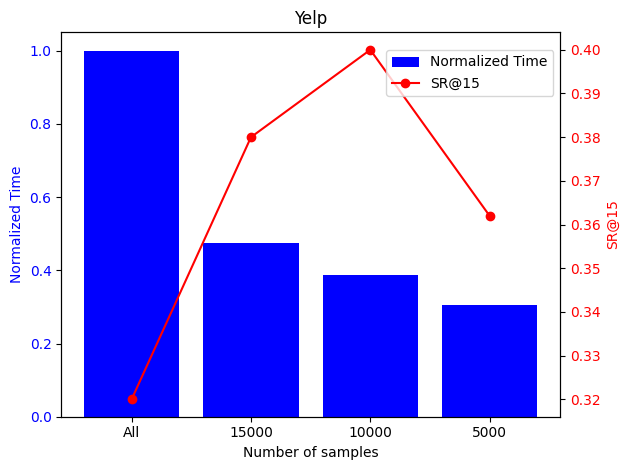}
    \label{f1}
\end{minipage}%
    \hfill%
\begin{minipage}[t]{0.49\linewidth}
    \includegraphics[width=\linewidth]{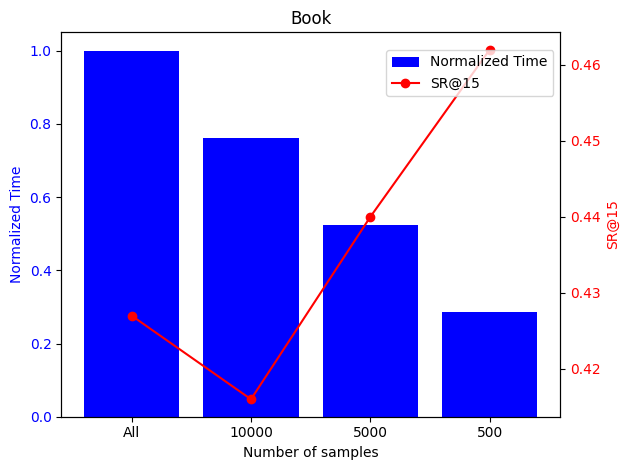} 
    \label{f2}
\end{minipage}
\caption{Impact of Node Sampling on normalized time cost and SR@15 performance.
} 
\label{fig:graph_sample_size}
\end{figure}

To understand the trade-off between computational efficiency and recommendation performance, we conducted experiments to analyze the impact of the graph sampling strategy introduced in Eq. \ref{eq:graph_node}. By reducing the number of item nodes in the dynamic conversation graph, the graph construction and node update process can be accelerated, potentially meeting the real-time requirements of conversational recommendation scenarios.

Figure \ref{fig:graph_sample_size} shows the results of our experiments on the Yelp and Amazon-Book datasets. The x-axis represents the number of sampled item nodes, and the y-axis shows the normalized time cost (compared to using the full candidate item set) on the left, and the corresponding Success Rate@15 (SR@15) on the right.

The graph-based conversation modeling approach, while computationally intensive, can be optimized for real-time scenarios through effective node sampling. The empirical evidence suggests that our sampling strategy not only reduces computational demands but also potentially enhances recommendation performance by focusing on a representative subset of items. 

The observed improvement in SR@15 with reduced sample sizes may be attributed to a more focused learning process and a higher signal-to-noise ratio in the node embeddings. Smaller, more relevant subsets of nodes allow the GCN to perform updates more effectively, leveraging the robust initial embeddings provided by TransE. This focused approach reduces the dilution of critical signals by extraneous data, enhancing the model's ability to discern and adapt to the most significant user-item interactions for recommendations. Additionally, computational efficiencies achieved by sampling facilitate more frequent updates, allowing the model to adapt more dynamically to evolving user preferences within the conversational context.

\subsection{Case Study (RQ6)}
\begin{figure*}[t]
    \centering
	\includegraphics[width=0.7\linewidth]{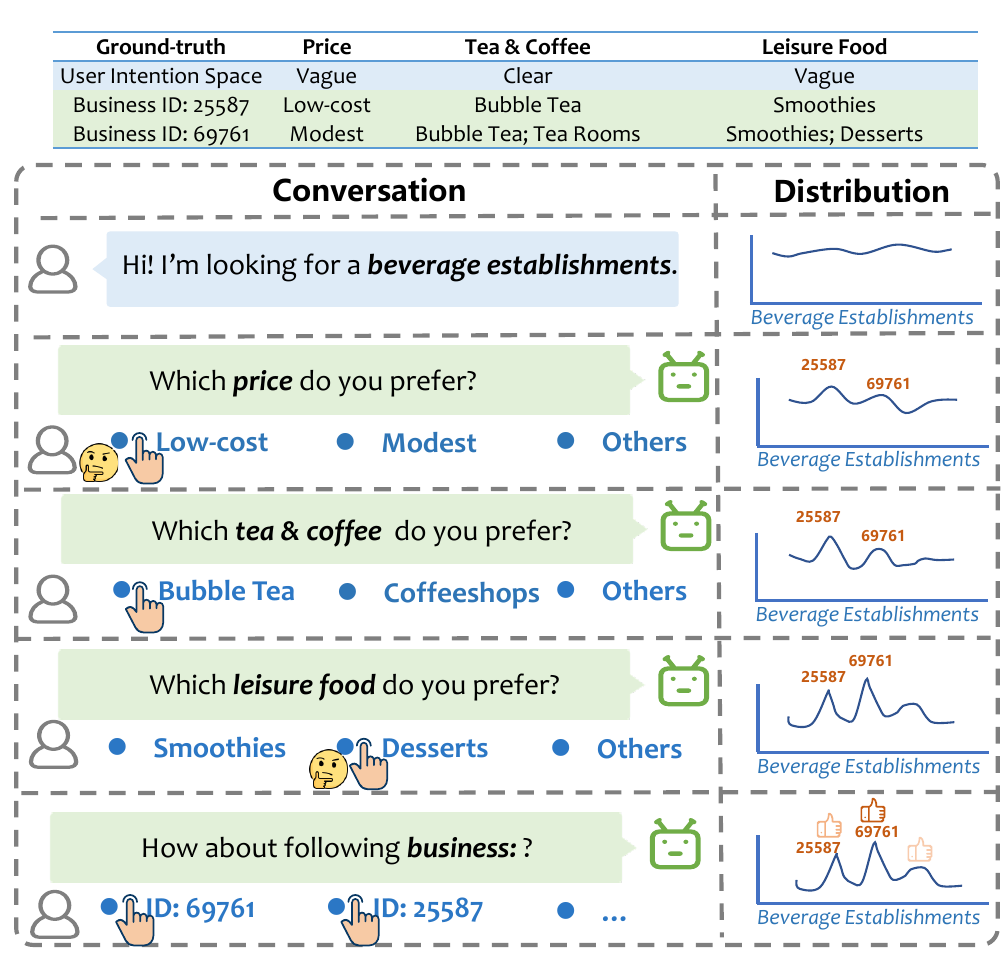}
    \caption{The left figure is a sample conversation generated by our VPPL and the right one shows changes in the user preference distribution during the conversation. The top of the figure displays the user’s clear preference and vague preferences toward target items in the VPMCR scenario.}
    \label{fig:case_study}
\end{figure*}

In this case study, we illustrate an example from the Yelp dataset (see Figure \ref{fig:case_study}). The user initiates the conversation with a clear preference for a beverage shop, which defines the user’s distribution space over all relevant items. In this VPMCR scenario, the user has a clear preference for “tea $\&$ coffee” but vague preferences for “price” and “leisure food”. The user will explicitly choose “Bubble Tea” in the second round, while they may randomly choose “low-cost” price and “dessert” food in the first and third rounds, respectively.

Our method, VPPL, effectively captures the user’s vague and dynamic preferences by incorporating both directed preferences and inferred preferences from the user’s click/non-click behavior. It updates the user’s preference distribution over all candidate items using a soft estimation mechanism, which avoids the over-filtering problem and maintains a diverse and accurate item space. Even if the user does not explicitly click on some potentially preferred attributes in the early rounds, resulting in a lower confidence score for some items, these items still have a chance to be reactivated in later rounds, eventually leading to the recommendation of the target item. For example, in the first round, because the user does not select the “modest” price, the preference of item “ID 69761” decreases. However, after the third round, the user’s current feedback (clicking on “dessert”, not clicking on “smoothies”) and historical feedback (“price” and “tea $\&$ coffee”) combine to produce two items with the highest preference scores, namely “ID 69761” and “ID 25587”.

Finally, our method successfully recommends two satisfactory items that match the user’s multi-interest preferences. This case study demonstrates the advantages of VPPL in handling vague user preferences in CRS.

\section{Conclusion}
In this paper, we introduce a novel and realistic scenario for conversational recommendation systems (CRSs), called Vague Preference Multi-round Conversational Recommendation (VPMCR). VPMCR aims to address the inherent ambiguity and relative decision-making processes exhibited by users, who often exhibit vague preferences without well-defined inclinations for certain attribute types (e.g., color, pattern). To effectively navigate the VPMCR scenario, we proposed the Vague Preference Policy Learning (VPPL) framework, a unified approach that employs two main components: Ambiguity-aware Soft Estimation (ASE) and Dynamism-aware Policy Learning (DPL). 

ASE accommodates user preference ambiguity by calculating preference scores for both directed and inferred preferences, utilizing a choice-based method. This method extracts preferences from user interactions—both from selections and non-selections.  It also incorporates a time-aware preference decay strategy to model the dynamic nature of user preferences, giving more weight to recent preferences while gradually diminishing the influence of historical preferences.

DPL implements a policy learning framework, leveraging the preference distribution from ASE to guide the conversation and adapt to changes in users' preferences for making recommendations or querying attributes. It models the conversation using a dynamic heterogeneous graph, with edge weights reflecting ASE's soft estimation scores. To enhance graph modeling and policy learning efficiency, DPL introduces a novel graph sampling strategy and a preference-guided action pruning technique, optimized through reinforcement learning.

The VPPL framework within the VPMCR scenario represents a significant advancement in accommodating the complexities of user preferences and decision-making in CRS. Through comprehensive experiments conducted on four diverse datasets, we demonstrated the superior performance of VPPL, outperforming existing methods and setting a new benchmark for CRS research. Our work signifies a pivotal shift from conventional binary feedback models, emphasizing the fluidity and subtlety inherent in user decision-making processes in CRS environments.

While our work addresses some key limitations of existing CRS methods, we acknowledge that our assumptions about users' vague preferences may not be sufficiently refined. In future research, we plan to conduct more detailed modeling of user profiles to construct a more realistic representation of users' vague preferences in conversational settings. In particular, we aim to better account for shifts in user interests during conversations, as these are potential areas for further exploration and improvement.

\section*{Acknowledgements}
This work is supported by the National Natural Science Foundation of China (62402470), the Fundamental Research Funds for the Central Universities of China (WK2100000053, PA2024GDSK0107), Anhui Provincial Natural Science Foundation (2408085QF189), and the Postdoctoral Fellowship Program of CPSF (GZC20241643). This research is supported by the advanced computing resources provided by the Supercomputing Center of the USTC.


\bibliographystyle{ACM-Reference-Format}
\bibliography{reference}

\end{document}